\documentclass[a4paper,fleqn,usenatbib]{mnras}
\usepackage{graphicx}
\usepackage{amssymb}
\usepackage{amsmath}
\usepackage{newtxtext,newtxmath}
\usepackage[T1]{fontenc}
\usepackage{ae,aecompl}
\usepackage{color}

\title[Virgo Cluster Outskirts]{Witnessing the Growth of the Nearest Galaxy Cluster: Thermodynamics of the Virgo Cluster Outskirts}

\author[A. Simionescu et al.]{
A. Simionescu,$^{1}$\thanks{E-mail: aurora@astro.isas.jaxa.jp}
N. Werner,$^{2,3}$
A. Mantz,$^{4,5}$
S. W. Allen,$^{4,5,6}$
O. Urban$^{4,5}$
\\
$^{1}$Institute of Space and Astronautical Science (ISAS), JAXA, 3-1-1 Yoshinodai, Chuo-ku, Sagamihara, Kanagawa, 252-5210 Japan\\ 
$^{2}$MTA-E\"otv\"os Lor\'and University Lend\"ulet Hot Universe Research Group, H-1117 P\'azm\'any P\'eter s\'et\'any 1/A, Budapest, Hungary\\ 
$^{3}$Department of Theoretical Physics and Astrophysics, Faculty of Science, Masaryk University, Kotlarsk\'a 2, 611 37 Brno, Czech Republic\\
$^{4}$KIPAC, Stanford University, 452 Lomita Mall, Stanford, CA 94305, USA\\
$^{5}$Department of Physics, Stanford University, 382 Via Pueblo Mall, Stanford, CA  94305-4060, USA\\
$^{6}$SLAC National Accelerator Laboratory, 2575 Sand Hill Road, Menlo Park, CA 94025, USA\\
}

\date{Accepted XXX. Received YYY; in original form ZZZ}

\pubyear{2016}

\begin{document}
\label{firstpage}
\pagerange{\pageref{firstpage}--\pageref{lastpage}}
\maketitle

\begin{abstract}
We present results from Suzaku Key Project observations of the Virgo Cluster, the nearest galaxy cluster to us, mapping its X-ray properties along four long `arms' extending beyond the virial radius. 
The entropy profiles along all four azimuths increase with radius, then level out beyond $\sim$0.5$r_{200}$, while the average pressure at large radii exceeds Planck Sunyaev-Zel'dovich measurements. These results can be explained by enhanced gas density fluctuations (clumping) in the cluster's outskirts.
Using a standard Navarro, Frenk and White (1997) model, we estimate a virial mass, radius, and concentration parameter of $M_{200}=1.05\pm0.02\times10^{14}$~M$_\odot$, $r_{200}=974.1\pm5.7$~kpc, and $c = 8.8 \pm0.2$, respectively. The inferred cumulative baryon fraction exceeds the cosmic mean at $r\sim r_{200}$ along the major axis, suggesting enhanced gas clumping possibly sourced by a candidate large-scale structure filament along the north-south direction.
The Suzaku data reveal a large-scale sloshing pattern, with two new cold fronts detected at radii of 233~kpc and 280~kpc along the western and southern arms, respectively. Two high-temperature regions are also identified 1~Mpc towards the south and 605~kpc towards the west of M87, likely representing shocks associated with the ongoing cluster growth. 
Although systematic uncertainties in measuring the metallicity for low temperature plasma remain, the data at large radii appear consistent with a uniform metal distribution on scales of $\sim 90\times180$~kpc and larger, providing additional support for the early chemical enrichment scenario driven by galactic winds at redshifts of 2--3.
\end{abstract}


\section{Introduction}\label{intro}

Structure in the Universe on the largest scales evolved via the gravitational collapse of initially small density perturbations. Clusters of galaxies are the largest virialized knots of the cosmic web, and are still growing at the present time via mergers and accretion from the surrounding large-scale structure filaments. As such, the outer edges of clusters of galaxies, the regions that mark the boundary between the virialized and in-falling material, provide important laboratories for studying the formation and evolution of the cosmic web. Most of the baryonic mass in clusters of galaxies is in the form of a tenuous, hot plasma that emits primarily at X-ray wavelengths; a reliable characterisation of the thermodynamical properties of the very faint, diffuse gas in the outskirts of these systems has been particularly challenging in that it has required very careful modelling of both the instrumental and cosmic X-ray backgrounds and foregrounds. A significant investment in both observing time and dedicated data analysis efforts over the last years has led to important progress in our understanding of the regions of ongoing virialization that bear witness to the most recent processes related to large-scale structure growth (for a review, see \citealt{reiprich2013} and references therein, as well as more recent results by e.g. \citealt{urban2014,sato2014,su2015,eckert2015}). In addition to X-ray observations, the imprint of the hot, diffuse intra-cluster medium (ICM) onto the cosmic microwave background (CMB) through the Sunyaev-Zel'dovich (SZ) effect can also be used to extend our knowledge of the thermodynamical properties (in particular the plasma pressure in the case of SZ measurements) out to the virial radii of galaxy clusters and into the cosmic web filaments beyond \citep[e.g.][]{planck2013}.

Here, we present the results from a Suzaku Key Project dedicated to mapping the outskirts of the Virgo Cluster in X-rays along several azimuths using more than 60 different pointings and over a megasecond of total net exposure time. At a distance estimated between 15 and 21 Mpc \citep{pierce1994,federspiel1998,mei2007}, Virgo is the nearest cluster to our Galaxy, and the largest such system in terms of angular size, enabling us to look for substructure in the ICM in unprecedented spatial detail. It is moderate in terms of its average X-ray temperature ($2.4^{+0.3}_{-0.2}$ keV, \citealt{boehringer1994}), and its total mass is therefore similar to that of a large fraction of the tens-to-hundreds of thousands of galaxy clusters expected to be discovered by ongoing and future surveys, such as the Dark Energy Survey and eROSITA. A precise understanding of the growth and thermodynamic properties of clusters in this mass range will be critical to the extraction of robust cosmological parameters.

As the nearest and second-brightest extended extragalactic soft X-ray source, the Virgo Cluster has been observed numerous times over the years with previous and current generations of space telescopes. \citet{boehringer1994} used ROSAT Position Sensitive Proportional Counter (PSPC) observations from the ROSAT All-Sky Survey to study the structure of the Virgo Cluster and discovered that the X-ray emission in general traces the galaxy distribution reported by \citet{binggeli1993}, confirming 3 sub-clumps centered on the galaxies M87, M49, and M86 respectively. \citet{shibata2001} derived an extensive temperature map using hardness ratio values from ASCA observations covering an area of 19 deg$^2$, a large fraction of which is located south of the cluster centre, in the region connecting the M87 and M49 haloes. \citet{urban2011} studied for the first time the spectroscopically derived properties in the outskirts of the Virgo Cluster using a mosaic of XMM-Newton pointings, which cover it from the center northwards out to beyond its virial radius. 

In this paper, we take advantage of the relatively low and stable instrumental background of the Suzaku X-ray satellite, as well as the extensive data set available through the Virgo Cluster Key Project, to expand upon these previous studies and present the most detailed spectroscopic X-ray measurements of the thermodynamical properties in the intracluster medium of this system obtained to date out to its virial radius. Our results are complementary to the recent study presented in \citet{planck_virgo}, who used the SZ effect to map the entire Virgo Cluster up to twice its virial radius.

We assume a $\Lambda$CDM cosmology with $\Omega_{\rm m}$=0.27, $\Omega_\Lambda$=0.73, and H$_0$=70 km/s/Mpc. All errors are given at $1\sigma$ (68\%) confidence level unless otherwise stated. We assumed the distance of Virgo as that of the central galaxy M87 (16.1 Mpc, \citealt{tonry2001}). At this distance, 1 arcmin corresponds to 4.65 kpc. For consistency with previous cluster outskirts studies, throughout the manuscript, ``virial radius'' refers to $r_{200}$, the radius within which the mean enclosed matter density is 200 times the critical density at the redshift of the source.

\section{Observations and data reduction}

A large mosaic of observations of the Virgo Cluster was obtained as a Suzaku Key Project during AO-7 and AO-8 (PI: A. Simionescu). In addition, we use archival observations centred on M87, the brightest cluster galaxy (BCG), and on M49, a member galaxy located about 1.2~Mpc south of the cluster center, as well as several Suzaku pointings aligned along the northern arm and observed in previous cycles: two fields located immediately north of M87 (AO-3, PI: A. Simionescu) and seven fields covering the outer parts of the cluster, between 1.0 and 1.5~Mpc (AO-6, PI: N. Werner). The details of all the observations used in this work are summarised in Table \ref{tab:observations}. 

Initial results from this data set were presented by \citet{Simionescu2015}, who focused on the distribution of chemical elements in the intra-cluster medium on very large scales. Here, we present  the detailed thermodynamic properties of the intra-cluster medium up to the virial radius of the Virgo Cluster.

\begin{figure*}
\begin{center}
\includegraphics[width=0.85\textwidth]{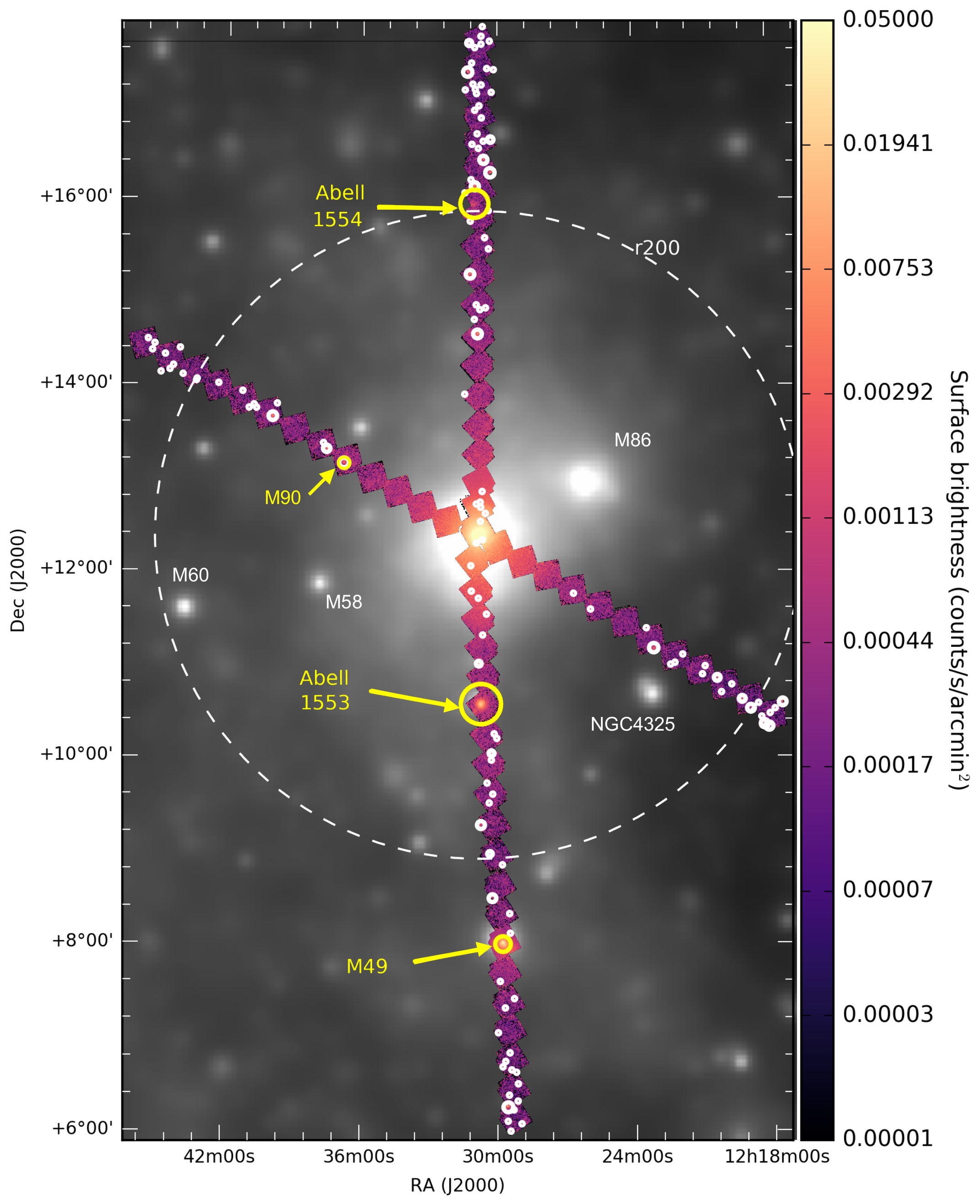}
\caption{An image of the complete Suzaku mosaic of the Virgo Cluster in the 0.7-2.0 keV energy band, corrected for instrumental background and vignetting, is shown in false colour. Point sources excluded from further analysis are shown as white circles, while exclusion regions around several known extended objects are shown in yellow and labeled. The approximate value of $r_{200}$ for the Virgo Cluster is shown as a dashed white circle. For the regions not covered by Suzaku, the corresponding ROSAT All-Sky Survey map is shown in grayscale, and several additional substructures not covered by the present analysis are labeled in white.}\label{img}
\end{center}
\end{figure*}

To reduce the data from Suzaku's X-ray Imaging Spectrometers (XIS) 0, 1 and 3, we follow the same analysis methods laid out in \citet{Simionescu2015}.
In brief, we applied the screening criteria recommended by the instrument team\footnote{Arida, M., XIS Data Analysis, http://heasarc.gsfc.nasa.gov/docs/suzaku/analysis/abc/node9.html (2010)}, filtered the data to include observation periods with the geomagnetic cut-off rigidity COR $>$ 6 GV, and excluded two columns on either side of the charge-injected columns in XIS1 to avoid charge leak known to affect these areas of the detector. Furthermore, we used a mask file provided by the XIS team to manually filter out pixels that are known to be hot or flickering\footnote{http://www.astro.isas.ac.jp/suzaku/analysis/xis/nxb\_new (2015)}. We applied the contamination layer calibration from 2014 August 25. The instrumental background was determined from night Earth observations in the standard way, using the task \texttt{xisnxbgen} \citep{tawa2008}.

\section{X-ray Surface Brightness}\label{sect_sx}

We combined images from the three XIS detectors to obtain an X-ray surface brightness map from the complete Suzaku mosaic in the 0.7--2.0 keV energy band, corrected for instrumental background and vignetting. The vignetting effect was accounted for by creating flat-field maps using the task {\tt xissim}, which performs a ray-tracing Monte Carlo simulation of the X-ray telescope (XRT) and the XIS detector response\footnote{see https://heasarc.gsfc.nasa.gov/docs/suzaku/analysis/expomap.html}. 
The resulting surface brightness distribution is shown in Figure \ref{img}.

Point sources were identified based on this image using the CIAO tool \texttt{wavdetect} employing a single wavelet radius of 1 arcmin, which is matched to the Suzaku half-power radius. These regions were then excluded from further analysis, together with three additional larger circular regions corresponding to M49 and to the background clusters Abell 1553 and Abell 1554 (with radii of 5, 13, and 9 arcmin, respectively; see Fig \ref{img}). \citet{MSPM_cat} have used an algorithmic search to identify groups, clusters, and large-scale structure filaments based on Sloan Digital Sky Survey data, including the regions covered by our Virgo Cluster mosaic. We have chosen not to excise the spatial regions corresponding to structures identified in this catalog, but have verified that this does not affect the measurements presented below.

\begin{figure*}
\begin{center}
\includegraphics[width=0.9\textwidth]{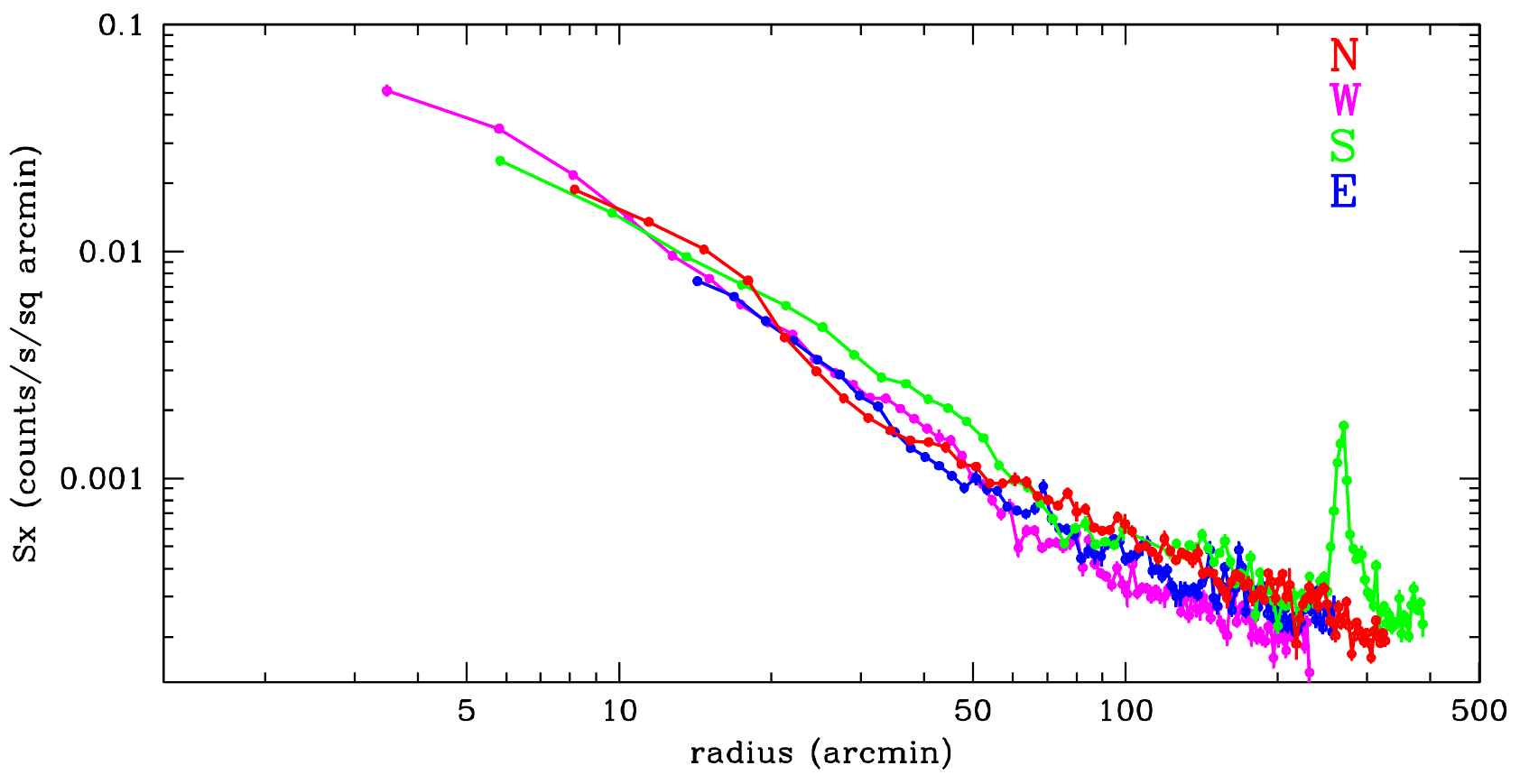}
\caption{Surface brightness profiles along the four arms in the Suzaku mosaic, using the 0.7-2 keV energy band. The profiles have been corrected for the instrumental background and for vignetting effects.}
\label{sb}
\end{center}
\end{figure*}

The surface brightness profiles obtained along the four arms in the Suzaku mosaic using the 0.7--2 keV energy band are plotted in Figure \ref{sb}. 
The cold front discussed by \citet{simionescu2010} is seen as a sharp gradient in the northern surface brightness at a radius of about 20 arcmin ($\sim$90~kpc). This is accompanied by similar features located further out along the other arms, most notably around 50 arcmin ($\sim$230~kpc) to the west and 60 arcmin ($\sim$280~kpc) towards the south.
The intra-cluster medium of the in-falling galaxy group associated with M49 is clearly seen as a surface brightness excess centred around a radius of 265 arcmin ($\sim$1.2~Mpc) along the southern arm. A clear asymmetry is observed at large radii, with the surface brightness along the E-W axis clearly smaller than along the N-S direction. In particular along the western arm the surface brightness falls off much faster as a function of radius compared to the other azimuths covered by the Suzaku mosaic.

\section{Spectroscopic mapping}

For each arm, we then extracted spectra from partial annuli centered on the core of M87. We imposed a minimum of 1200 net counts per spectral extraction region, accounting for the instrumental and sky backgrounds, and a minimum width of each annulus of 3 arcmin, taking into consideration the point-spread function (PSF) of Suzaku's X-ray telescopes.

The spectral modelling was performed with the XSPEC spectral fitting package \citep[version 12.8.2,][]{arnaud1996}, employing the modified C-statistic estimator. The spectra were binned to a minimum of one count per channel and fitted in the 0.7--7.0 keV band. The ICM was modelled as a thermal plasma in collisional ionization equilibrium using the {\it apec} code \citep{smith2001}. The gas metallicity is reported with respect to the proto-solar units of \cite{lodders2003}. The abundance ratios between different chemical elements were fixed at Solar, which was shown by \citet{Simionescu2015} to hold throughout the entire volume of the Virgo Cluster. The Galactic absorption column density, $n_H$, was fixed to the average value at the location of each respective Suzaku field based on the radio HI survey of \cite{kalberla2005}. 

\subsection{Background Modeling}\label{sect_bkgmo}

The instrumental background was subtracted in the standard way, using the task \texttt{xisnxbgen} \citep{tawa2008}. The cosmic X-ray background model consisted of a power-law emission component that accounts for the unresolved population of point sources (CXB), one absorbed thermal plasma model for the Galactic halo (GH) emission, and an unabsorbed thermal plasma model for the Local Hot Bubble (LHB) \citep[see e.g.][]{kuntz2000}. 

The average flux and spatial variation of the GH and LHB emission were determined using a set of 12 ROSAT All-Sky Survey (RASS) spectra obtained from circular regions with a radius of 1 degree, as described in detail in Section 3.1 of \citet{Simionescu2015}. The metallicity of both components was assumed to be Solar. The best-fit model parameters are summarised in Table \ref{tab_cxb}.

With these assumptions for the GH and LHB model parameters, in combination with the expected CXB model from previous Suzaku studies (e.g. \citealt{urban2014}, \citealt{Simionescu2013}), we find no significant ICM emission in the outermost three Suzaku fields along the N, S, and W directions, and the outermost two pointings along the E arm. We have fitted the spectra obtained from the full field of view of these observations in the 0.7--7 keV band, with the normalisation of the CXB model allowed to vary for each pointing separately, the GH normalisations for pointings corresponding to the same arm linked to one common value, and the GH temperature and LHB temperature and normalisation fixed to the best-fit parameters determined from the RASS data. We  used the variance between the 11 different Suzaku background fields to estimate the systematic uncertainty in determining the CXB normalisation. As shown in Table \ref{tab_cxb}, we obtain a good agreement between the Suzaku and RASS estimates for the GH fluxes along each of the four azimuths. For the rest of the manuscript, we consider the RASS values for the GH normalisation as the default model. Further considerations of the systematics arising from this choice are discussed in Section \ref{sect_hf} of the Appendix.

\begin{table}
\caption{Background model parameters. The first set of errors gives statistical uncertainties at 68\% confidence; the second set represents systematic uncertainties due to spatial variations.}
\begin{center}
\begin{tabular}{ccccc}
\hline
\hline
 & kT (keV) & arm & RASS norm & Suzaku norm \\
 & or $\Gamma$ & & $\times 10^{-6}$/arcmin$^2$ & $\times 10^{-6}$/arcmin$^2$ \\
\hline 
CXB & 1.50$^{*}$ & all & & $0.89$  \\
 & & & & $\pm0.06\pm0.10$ \\ 
\hline
 & & N & $1.31\pm0.10\pm0.22 $ & $1.24\pm$0.09 \\
GH & $ 0.20$ & W & $1.14\pm0.10\pm0.12$ & $1.27\pm0.10$ \\
 &$\pm0.01$ & S & $2.20\pm0.15\pm0.66 $ & $2.33\pm0.14$ \\
 & & E & $1.73\pm0.12\pm0.11$ & $2.00\pm0.16$\\
\hline
 & & N & $1.08\pm0.04\pm0.08 $ & \\
LHB & $0.104$ & W & $1.35\pm0.05\pm0.05 $ & \\
 & $\pm0.002$ & S & $1.46\pm0.06\pm0.27 $ & \\
 & & E & $1.41\pm0.05\pm0.11$ & \\
\hline 
\end{tabular}
\end{center}
$^{*}${\footnotesize This parameter anti-correlates with the GH normalisations when performing the fit using the 0.7--7 keV energy band and was therefore kept fixed during the fit. We performed a separate fit in the 2--7 keV energy band where the GH contribution is negligible and found $\Gamma=1.53\pm0.03$, consistent with the assumed value for the full band fit.}
\label{tab_cxb}
\end{table}

\subsection{Projected Spectral Results}\label{sec_proj_prof}

\begin{figure*}
\begin{center}
\includegraphics[width=0.45\textwidth]{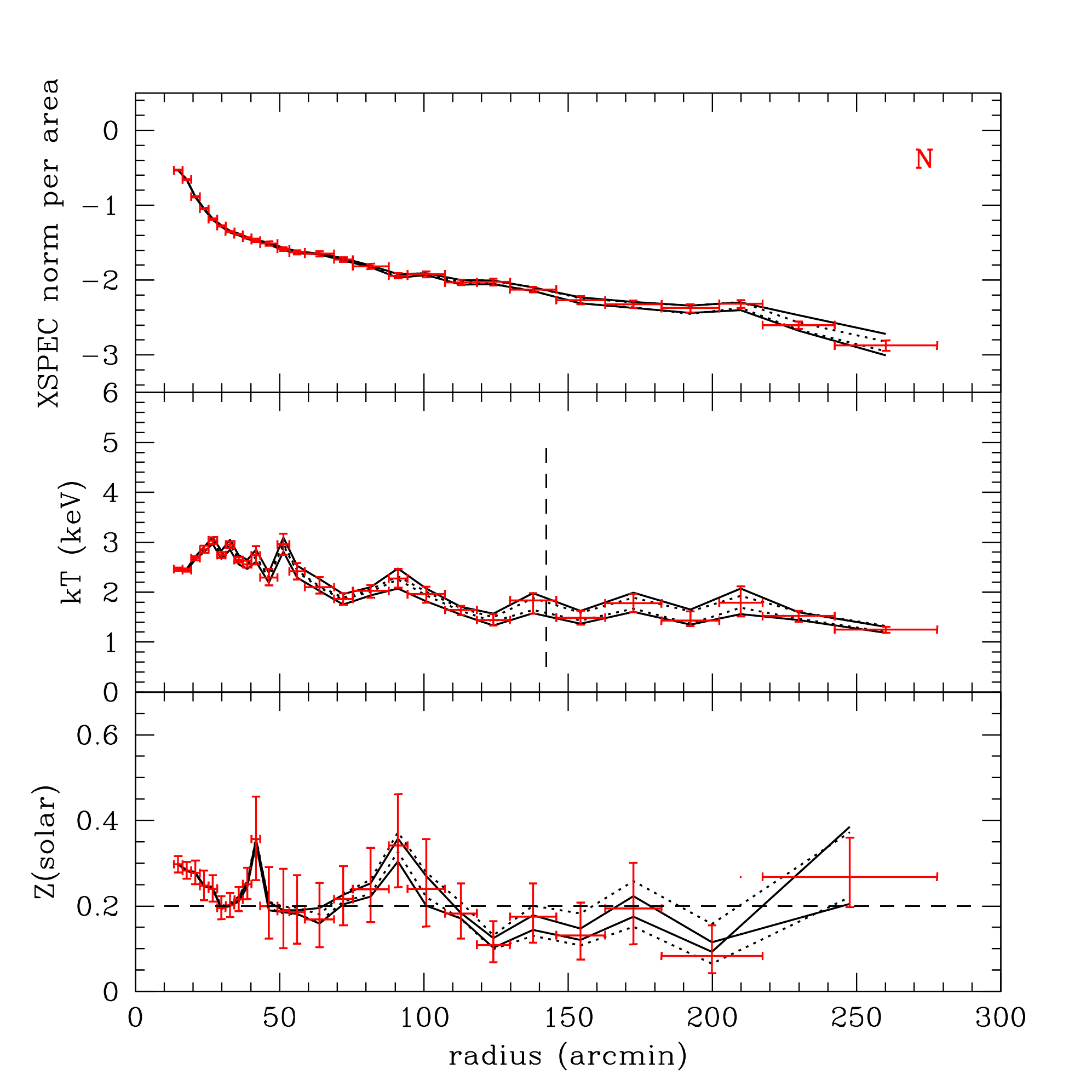}
\includegraphics[width=0.45\textwidth]{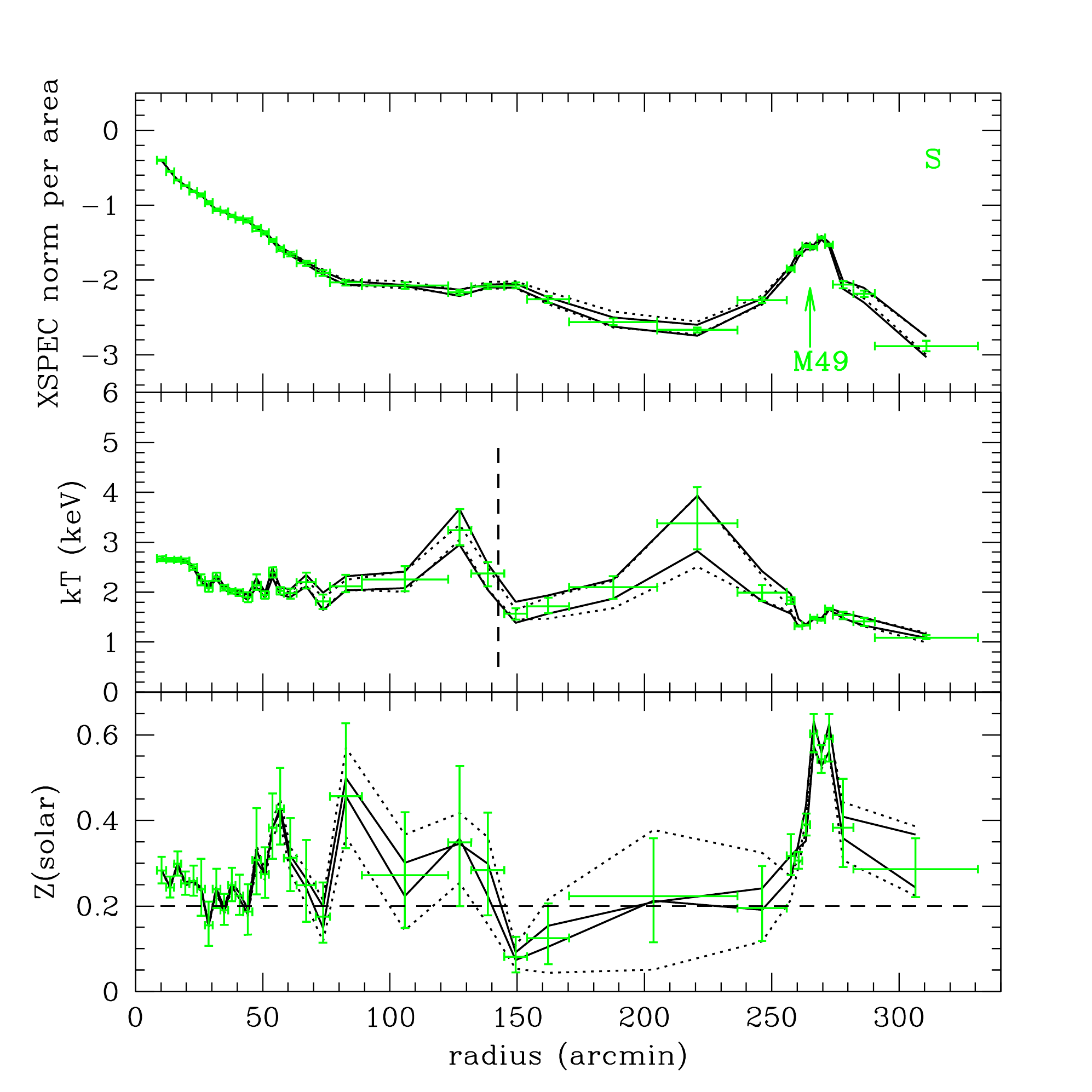}
\includegraphics[width=0.45\textwidth]{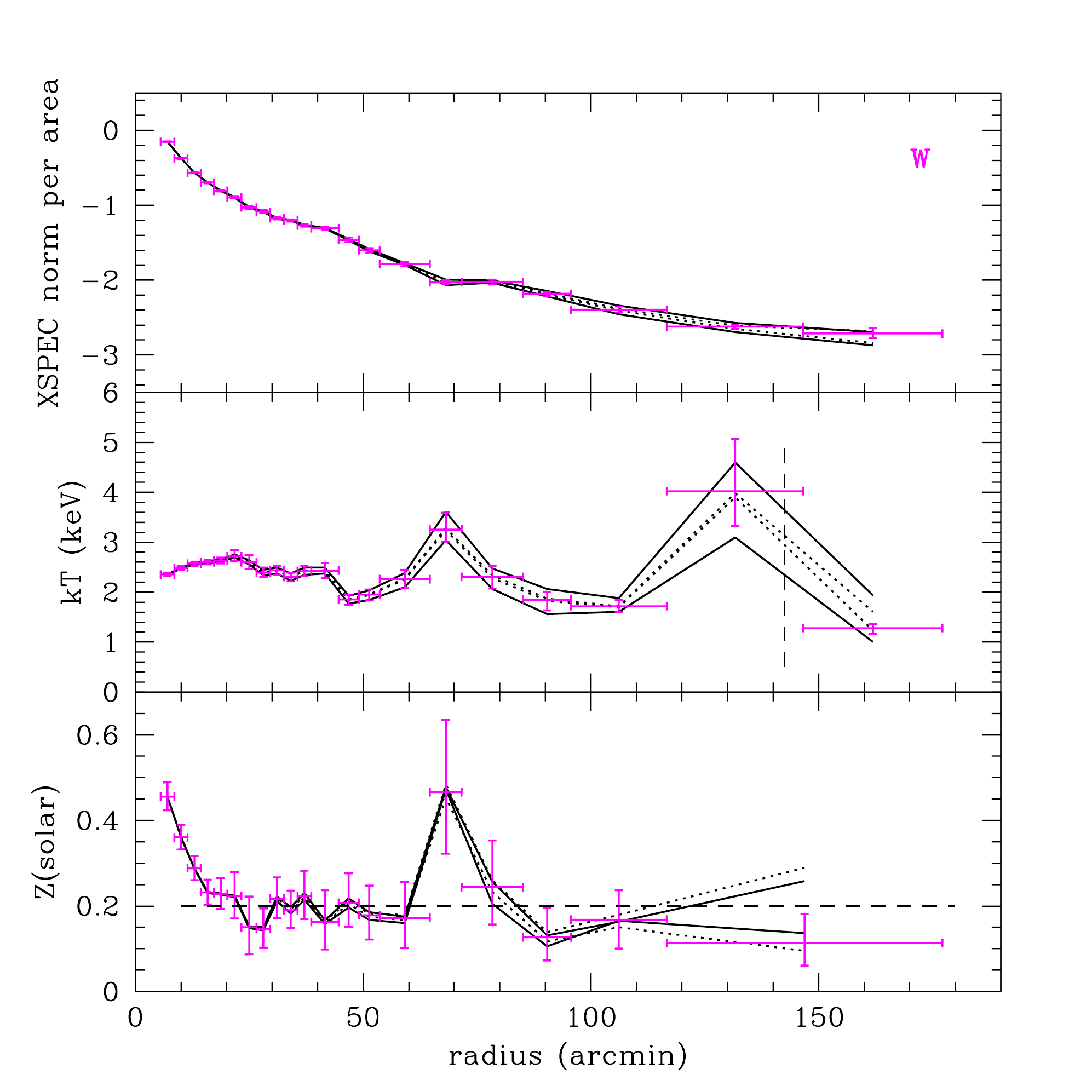}
\includegraphics[width=0.45\textwidth]{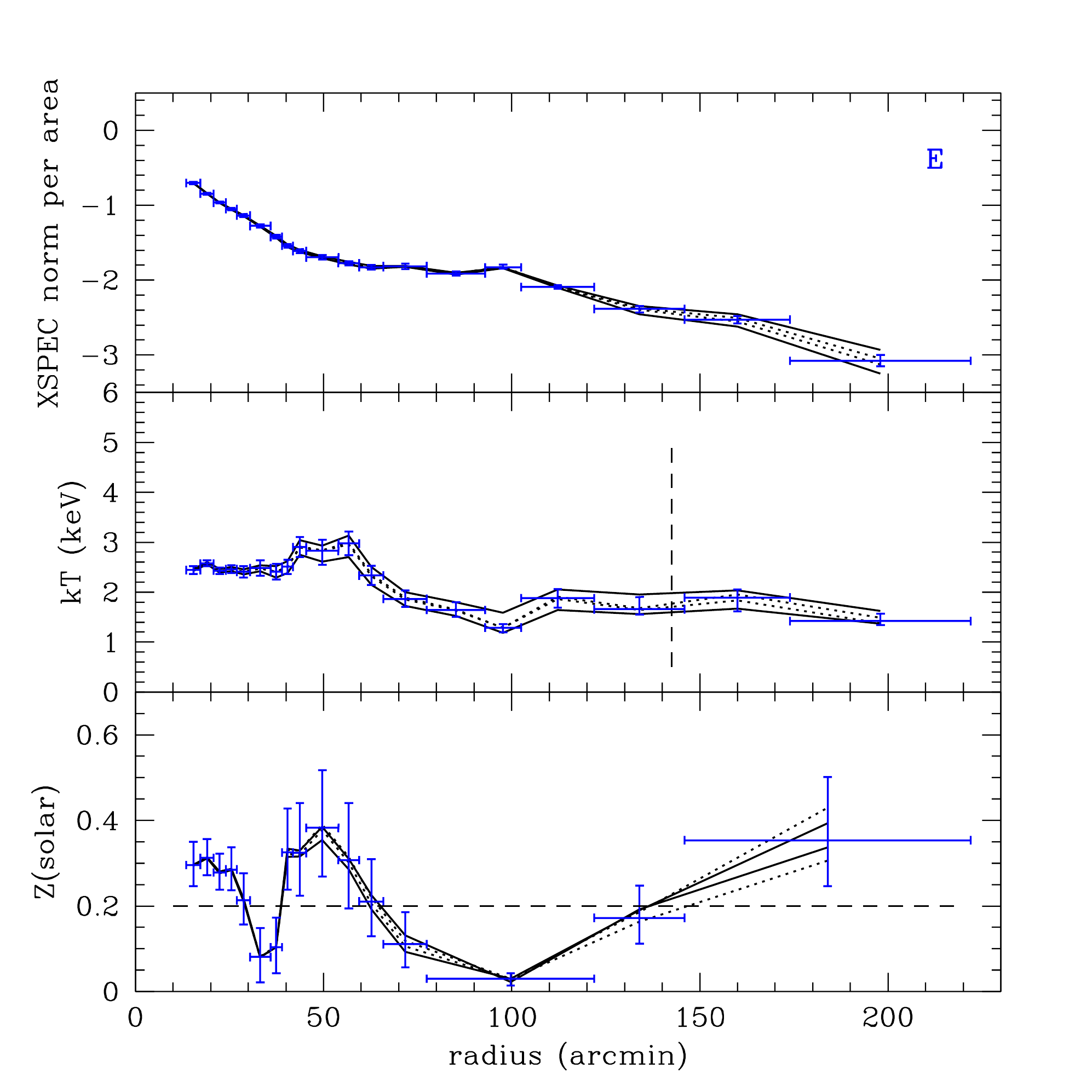}
\caption{Radial profiles of the projected spectrum normalisation, temperature, and metallicity along each of the four azimuths (red, top left: north; green, top right: south; magenta, bottom left: west; blue, bottom right: east). The effects of systematic uncertainties associated with the spatial variability of the GH and CXB flux are shown as dotted and solid black lines, respectively. The horizontal dashed black line in each bottom panel marks 0.2 Solar metallicity, while the vertical dashed black line in the middle panels shows the location of $r_{500}$.}\label{kT_Y_Z_proj}
\end{center}
\end{figure*}

In Figure \ref{kT_Y_Z_proj}, we show the best-fit radial profiles of the projected X-ray emission measure, temperature, and metallicity along the four arms in the Suzaku mosaic. In some cases, in order to constrain the metal abundance, neighbouring annuli were fitted in parallel with the temperature and spectrum normalisation free to vary independently but with a common metallicity. 

The systematic uncertainties due to the variation of the GH and CXB are also shown in Figure \ref{kT_Y_Z_proj}. In the case of the GH, we simply varied the normalisation of this model component within the ranges shown in Table \ref{tab_cxb} to determine its effects on the best-fit spectral properties in each annulus. The power-law normalisation in turn was changed by a factor proportional to the systematic error shown in Table \ref{tab_cxb} scaled by $\sqrt{\Omega_{\rm fov}/\Omega_{\rm sp}}$, where $\Omega_{\rm sp}$ is the area corresponding to each individual partial annulus. This is because the cosmic variance is expected to be directly proportional to $1/\sqrt{\Omega}$, while the value shown in Table \ref{tab_cxb} was calculated using extraction regions covering the size of a full XIS field of view (excluding a 30'' region around the detector edges), $\Omega_{\rm fov}=16.8\times16.8$ arcmin$^2$.  

\subsection{Thermodynamic Substructure}\label{sec_proj_feat}

Several noteworthy features can be seen in the projected profiles, and are robust to the systematic checks described above. 

Along the northern arm, beyond the prominent cold front at a 20 arcmin (90~kpc) radius discussed in detail by \citet{simionescu2010} and \cite{Werner2016}, the profiles appear relatively smooth, with the temperature remaining almost constant beyond about 100 arcmin.

Along the southern arm, between about 20 and 60 arcmin (90 -- 280 kpc), there is a clear surface brightness excess compared to the other azimuths (see also Figure \ref{sb}). This is associated with a temperature drop at 20 arcmin from the centre, beyond which the temperature remains relatively flat and starts to steadily increase again only beyond 70 arcmin. The southern surface brightness excess also appears to be correlated with an increase in gas metallicity, most prominent around 50--60 arcmin. We believe this feature to be associated with large-scale gas sloshing in the Virgo Cluster. Beyond this large-scale cold front, the southern temperature profile increases up to a local peak located at 125 arcmin, after which we see a second surface brightness excess centred around 150 arcmin. The origin of this feature is unclear, and may be related to adiabatic compression of the ICM due to the ongoing merger with the M49 group.

Further along the southern arm, we detect a high temperature region corresponding to the minimum surface brightness ``saddle'' point between M87 and the X-ray peak associated with the in falling group centred on M49. This high temperature region likely reflects the presence of compressed and shock heated gas at the interface between these two merging systems. The temperature increase corresponds to a factor of $1.70\pm0.34$ compared to the neighbouring annulus further out which, under the assumption of shock heating, would translate to a shock Mach number of $1.7\pm0.3$. Since it is difficult to estimate the density in this region before the start of the merger and the passage of the shock, we cannot verify whether the density compression would confirm or rule out such a Mach number. The M49 halo is seen as a clear surface brightness peak, and is associated with relatively low temperatures of around 1.2 keV. 

The western temperature profile reveals that the surface brightness edge seen at about 50 arcmin along this azimuth (as pointed out in Section \ref{sect_sx}) corresponds to a low-temperature region. This, along with the northern cold front at 20 arcmin and the southern cold front at 60 arcmin, is likely part of a large-scale low-temperature, high-density spiral pattern typical of large-scale sloshing seen in other clusters (see discussion in Sect. \ref{sect_sloshing}). As with the southern arm, the temperature increases beyond this western cold front up to a local maximum located at 70 arcmin and then decreases again towards the cluster outskirts. A second temperature peak is observed at about 130 arcmin (605~kpc) and may also be indicative of shock heating. The temperature jump of a factor of $3.14\pm0.71$ compared to the next annulus further out would indicate a shock Mach number of $2.7\pm0.4$. However, no significant corresponding density jump is seen in the X-ray emissivity profile. 

The temperature profile along the eastern direction shows a fairly gradual increase up to a local maximum around 50 arcmin and declines beyond this radius. Around a cluster-centric distance of about 100 arcmin, we observe an excess in X-ray emissivity associated with a dip in the temperature profile. This feature is most likely due to the presence of hot X-ray gas associated with the spiral galaxy M90 which may be a Virgo Cluster member. As for the northern arm, beyond 100 arcmin, the eastern temperature profile remains relatively flat out to the edge of the detection limit.

\subsection{Metal Abundance Distribution in the ICM}\label{sec_met}

The results presented by \citet{Simionescu2015} suggest a uniform chemical composition of the Virgo ICM, with the $\alpha$-element to Fe abundance ratios consistent with the Solar value out to the virial radius along all arms. However, that work necessarily relied on large integration regions in order to constrain the abundances of elements whose spectral emission lines are much weaker than Fe. Here, we investigate the metallicity profiles (assuming Solar abundance ratios between different elements) using a finer spatial grid.

\begin{figure}
\begin{center}
\includegraphics[width=0.45\textwidth]{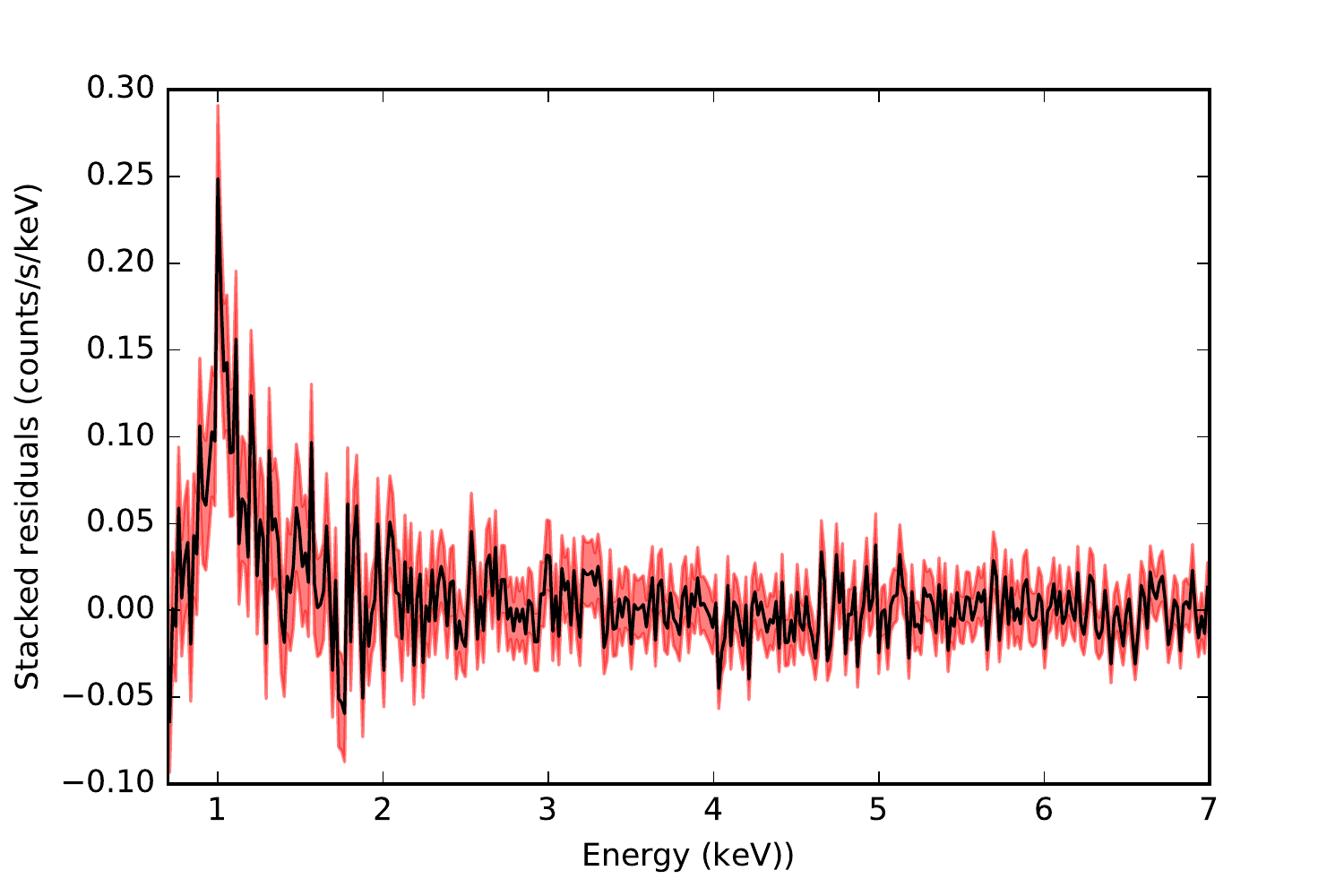}
\caption{Stacked residuals from the 15 spectra used to calculate the outermost metallicity data point along the N arm, with respect to the best-fit single temperature model in which the metallicity has been set to zero. The spectra have been binned by a factor of 5 for clarity. The black line shows the mean of the residuals, while the red band shows the statistical uncertainty on the measurement in each (binned) channel. The Fe-L emission is clearly detected from this region, located fully beyond $r_{200}$.}\label{stack}
\end{center}
\end{figure}

Remarkably, we are able to detect emission from the Fe-L complex out to beyond the cluster's $r_{200}$. Figure \ref{stack} shows the stacked residuals from all the spectra used to calculate the outermost metallicity data point along the N arm, upon having set the metallicity in the best-fit model to zero. This is the farthest region from the center of M87 where a metallicity measurement is reported, and spans 1.04--1.33 $r_{200}$. The residuals clearly show the Fe-L bump that constrains the abundance measurement when this is allowed to be non-zero in the fit.

The radial profiles of the metallicity along each of the four arms assuming a single-temperature model are shown in Figure \ref{kT_Y_Z_proj}. Prominent abundance peaks are observed at the locations of M87 and M49, as expected due to the ongoing production of metals by these massive elliptical galaxies. Apart from these regions, the metallicity values are distributed around an average of 0.2 Solar, with several noteworthy deviations listed below.

The southern arm shows an increased metallicity around 50--60 arcmin, which is most probably due to the large-scale sloshing transporting metal-enriched gas from the vicinity of the BCG out to larger radii (see \citealt{simionescu2010}). However, this metal excess extends even beyond the southern large-scale cold front, out to about 140 arcmin. 

Strong metallicity fluctuations are seen also along the eastern arm. In particular, the metallicity around 100 arcmin, coinciding with the location of M90, is unusually low, with a $3\sigma$ upper limit of only 0.07 Solar. This value does not increase significantly even when using a two-temperature model for the ICM emission (see Section \ref{sect_2T}), or when fixing the redshift in the spectral fit to the redshift of M90 rather than that of M87. The nature of the metallicity fluctuations along the eastern arm therefore remains unclear -- it is possible that a deteriorating gain stability of the XIS detectors in the late stages of the mission and/or an increase in the number of hot/flickering pixels that is not well calibrated may have an impact (the E arm observations were performed in AO-8 while all other data was collected during AO-7). 

Omitting the data within 150~kpc of M87, as well as all annuli located beyond the surface brightness minimum between M87 and M49, the southern cold front inside 60~arcmin, and the entire eastern arm that may be affected by the instrumental systematics mentioned above, we fitted a constant model to the metal abundance measurements and obtain an average value of $0.19\pm0.01$ Solar, with a $\chi^2=40.2$ for 39 d.o.f. suggesting that a uniform distribution already provides an overall acceptable fit. However, a linear model of the form $Z(r)=Z_0-r\times Z_r$ provides a significant improvement, with $Z_0=0.24\pm0.02$ Solar, $Z_r=0.12\pm0.04$ Solar/Mpc, and $\chi^2=30.6$ for 38 d.o.f. This would, at face value, suggest the presence of a radial gradient, with the average metallicity decreasing towards the cluster outskirts. The metal abundance measurements are driven by the strength of the Fe-L complex, which is known to be subject to strong biases in the presence of multi-temperature structure \citep{buote2000}; it is possible therefore that this radial gradient can be caused at least in part by the increased presence of multi-phase gas in the outskirts. This is discussed further in Section \ref{sect_2T}.

\section{Deprojection Analysis}\label{sect_deproj}

Assuming spherical symmetry in the individual arms, we have carried out a deprojection analysis using the XSPEC model \texttt{projct}. 
For this analysis, we fixed the ICM metallicity to 0.2 Solar in each annulus (see discussion in Sections \ref{sec_met} and \ref{sect_2T}). Given the wealth of substructure present in such a dynamically young system, it was necessary to combine annuli used in Section \ref{sec_proj_prof} until a monotonic emissivity and smoothly varying temperature profile were obtained, in order to ensure the stability of the fit. This reduces the spatial resolution of the deprojected profiles compared to the projected radial profiles, and several of the features discussed in Section \ref{sec_proj_feat} are washed out. Nonetheless, we caution the reader that the assumption of spherical symmetry remains crude. We further note that for the particular case of the southern arm, we have only used the data up until but excluding the high temperature region in the "saddle" point between M87 and M49.

This deprojection analysis does not account for emission from gas that lies beyond the outermost annulus, and which is projected onto this annulus. As a consequence, the density here tends to be overestimated (and, to a smaller extent, the density in the next annulus in is underestimated in turn). We have therefore fitted the deprojected density profile along each arm with a beta model and extrapolated this model to radii beyond our detection limit in order to correct for this effect. As a result, the corrected density in the outermost annulus along each azimuth becomes lower by a factor of between 0.23 (for the E arm) and 0.33 (for the W arm) compared to the initial best-fit values obtained directly from the spectral fit parameters. The density in the one-but-last annulus in turn increases only by less than 5\%. We note that no temperature correction was applied (which is equivalent to assuming that the ICM beyond the last annulus is isothermal and has the same temperature as this outermost region where a measurement can be obtained).

\begin{figure*}
\begin{center}
\includegraphics[width=0.45\textwidth]{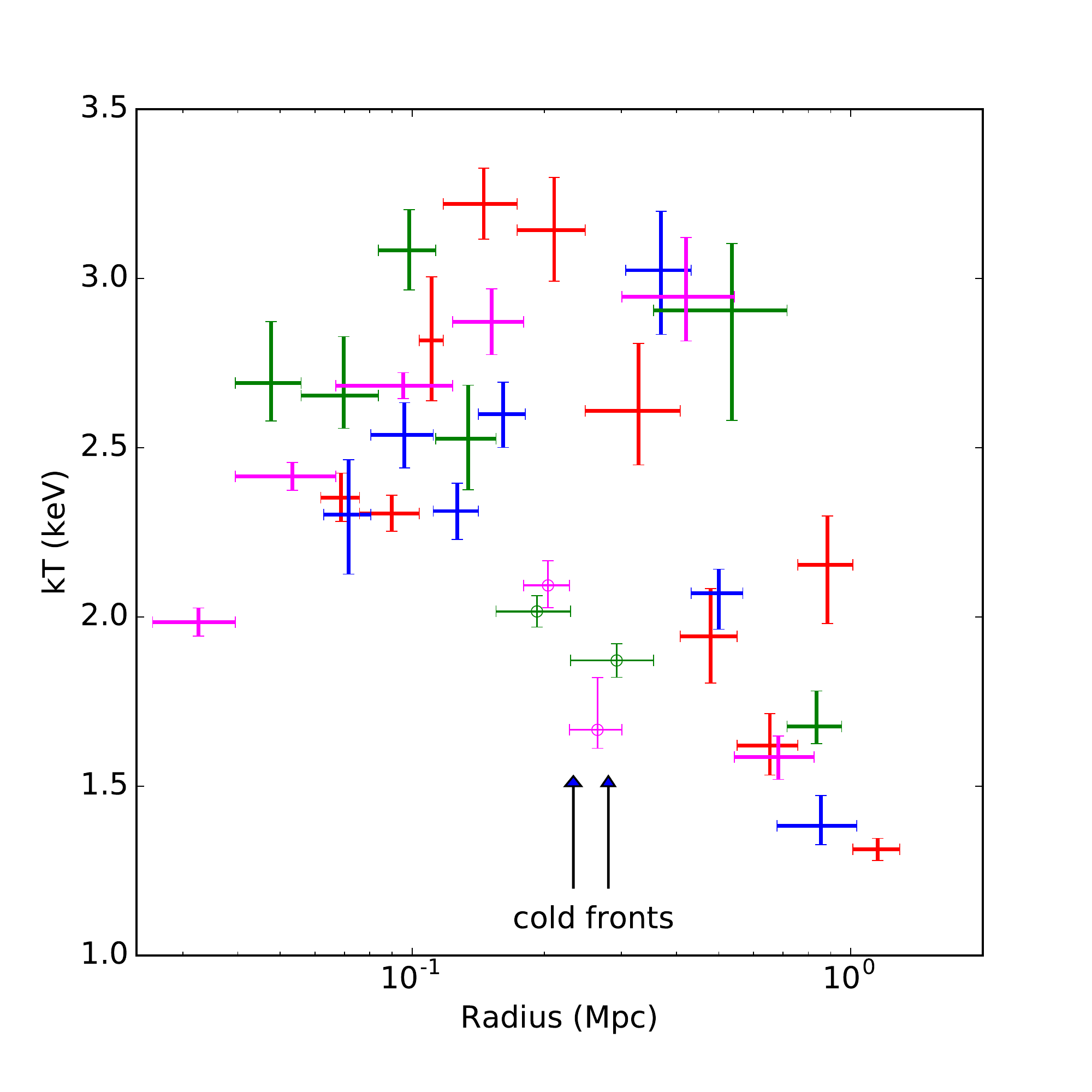}
\includegraphics[width=0.45\textwidth]{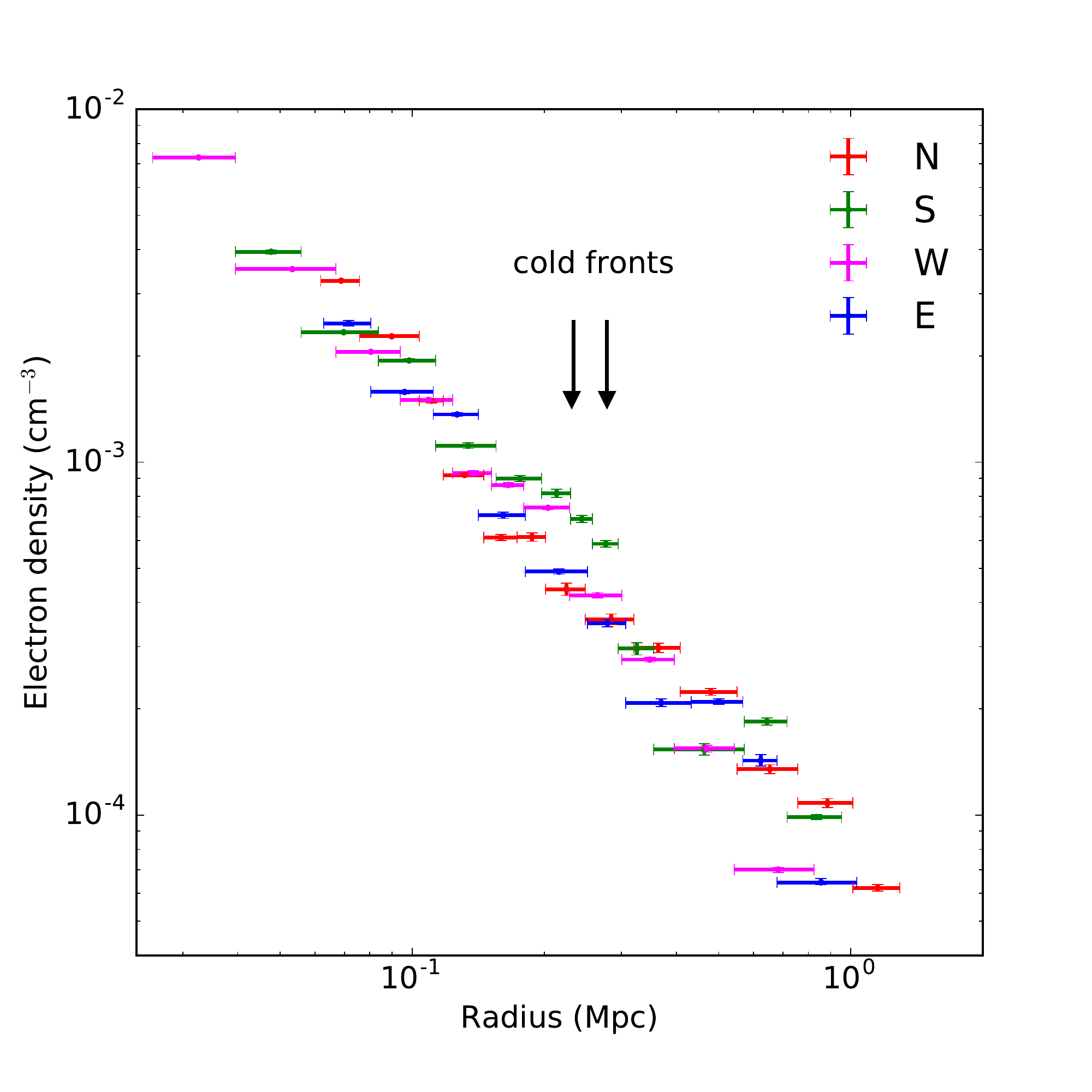}
\includegraphics[width=0.45\textwidth]{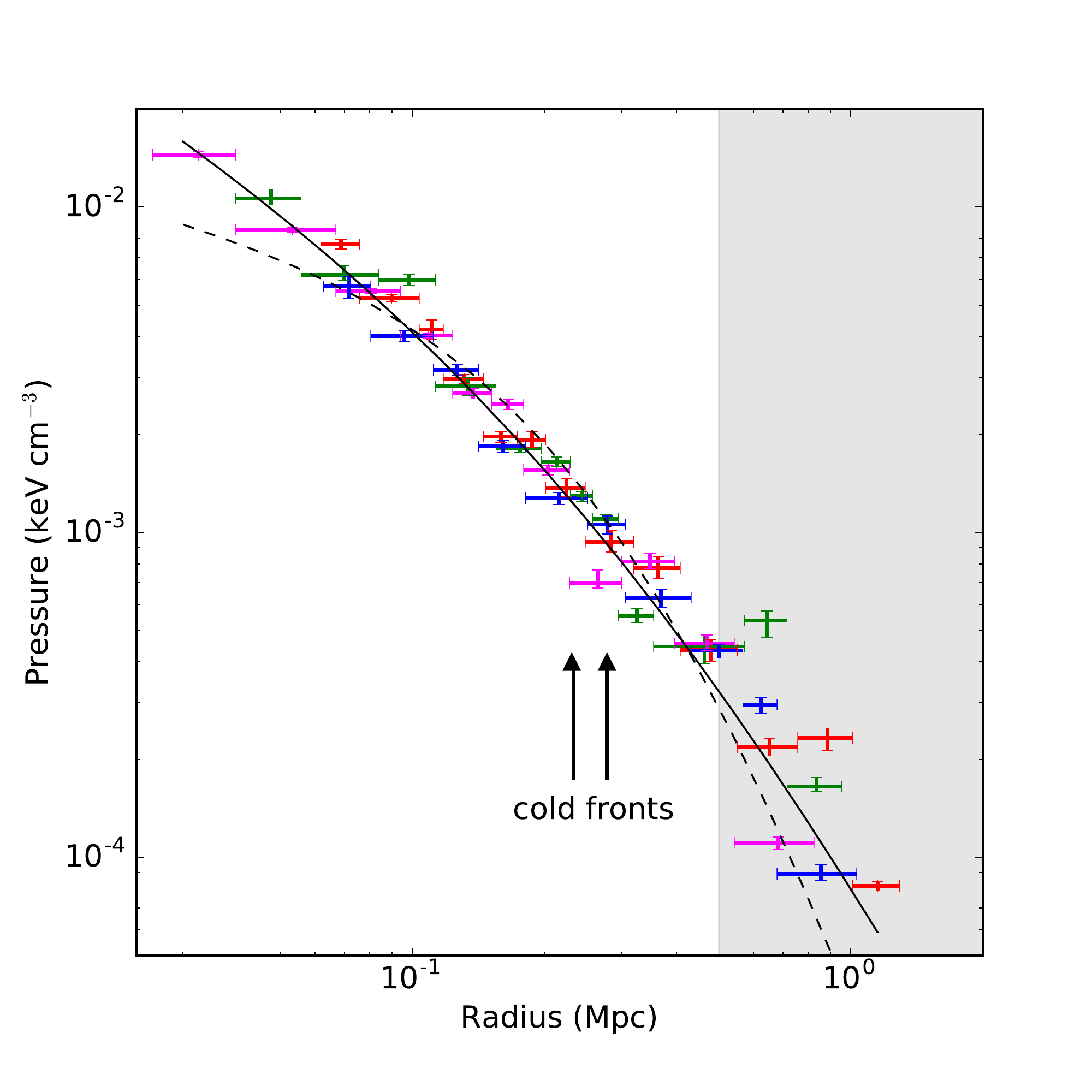}
\includegraphics[width=0.45\textwidth]{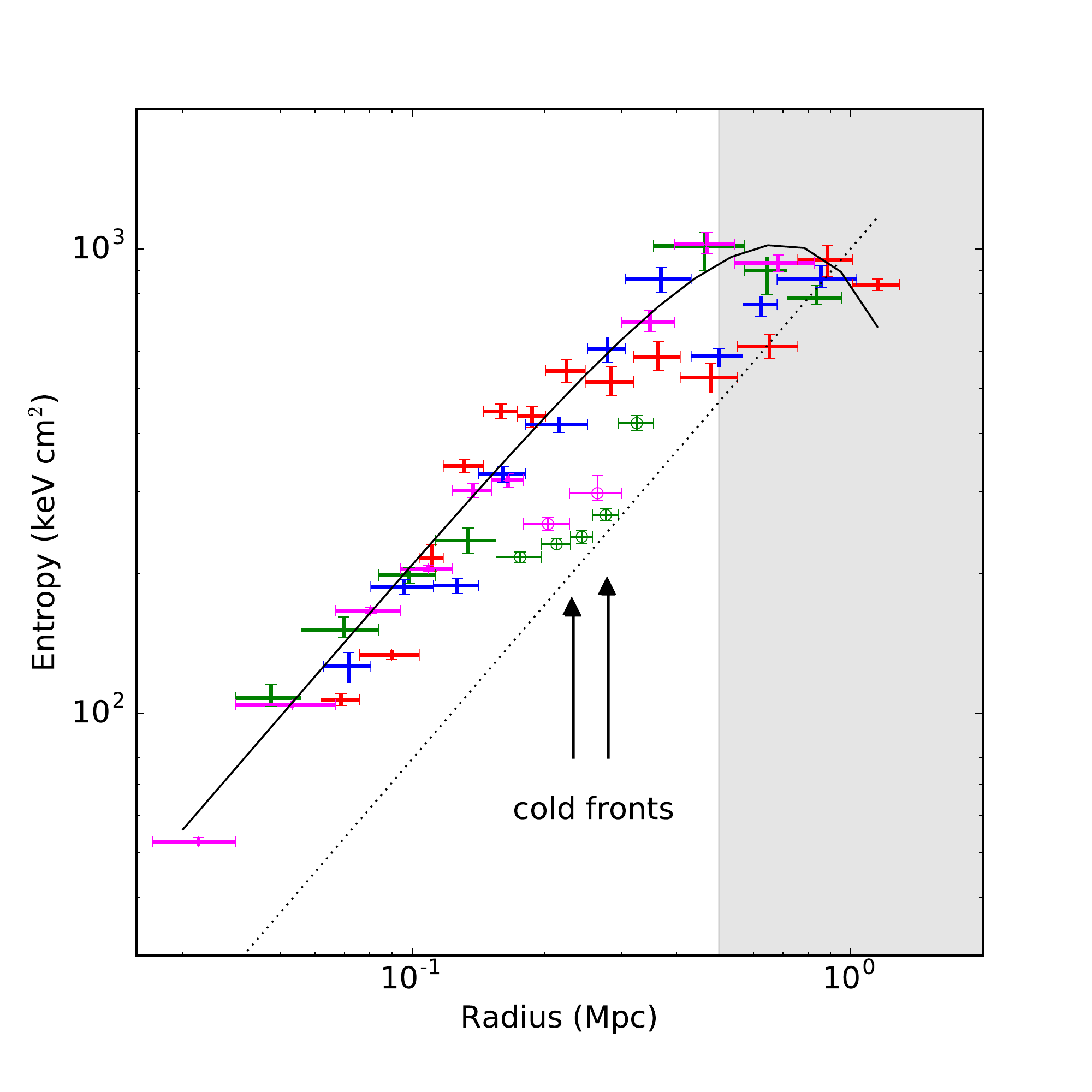}
\caption{Radial profiles of the deprojected thermodynamic properties of the Virgo Cluster ICM: temperature (top left), electron density (top right), pressure (bottom left) and entropy (bottom right). The western and southern cold fronts are shown with open symbols and thinner lines in the temperature and entropy panels. The base-line entropy model proposed by \citet{voit2005} for a baryon fraction $f_b=0.10$ is shown as a dotted black line, while the reference entropy model of \citet{walker2012c} is shown with a solid line. In the pressure panel, the average profile from a sample of 62 clusters observed by \citet{planck2013} is shown as a dashed line, while the best-fit generalised model based on Eqn \ref{eqn:pressure} is shown as a solid line. The region where we conclude that the entropy and pressure profiles become affected by gas clumping is marked with a grey background.}\label{deproj}
\end{center}
\end{figure*}

Figure \ref{deproj} shows the resulting profiles of deprojected temperature ($kT$), electron density ($n_e$), pressure ($P = n_ekT$), and entropy ($K = kT/n_e^{2/3}$), including the beta-model correction. The southern and western large-scale sloshing cold fronts are evident as low-temperature, high-density and low-entropy features in the deprojected thermodynamic profiles. At large radii, as expected from the projected surface brightness profiles, we find that the density drops off fastest along the western direction and generally the density along the minor (east-west) axis of the cluster is smaller than along the major axis running north-south.

\subsection{Pressure profiles}

In terms of the pressure profiles, \citet{nagai2007} propose a generalized model of the form
\begin{equation}
\frac{P(r)}{P_{500}}=\frac{P_0}{\left(c_{500}x\right)^\gamma\left[1+\left(c_{500}x\right)^\alpha\right]^{(\beta-\gamma)/\alpha}},
\label{eqn:pressure}
\end{equation}
where $P_{500}=1.45\times10^{-11}\,\text{erg}\,\text{cm}^{-3}\left(\frac{M_{500}}{10^{15}h^{-1}\,M_{\odot}}\right)^{2/3}E(z)^{8/3}$, $x=r/r_{500}$, $c_{500}$ is the concentration parameter defined at $r_{500}$, and the indices $\alpha$, $\beta$ and $\gamma$ are the profile slopes in the intermediate, outer and central regions, respectively. 
\citet{planck2013} studied the pressure profiles of 62 \emph{Planck} clusters between $0.02r_{500}<r<3r_{500}$, finding the best fit set of parameters $\left[P_0,c_{500},\alpha,\beta,\gamma\right]_{\text{Planck}}=\left[6.41,1.81,1.33,4.13,0.31\right]$. This model is shown as a dashed line in the lower left-hand panel of Figure \ref{deproj}, where we have used the values for $r_{500}$ and $M_{500}$ derived in Section \ref{sect:mass}. While it provides a reasonable description of the measurements at intermediate radii, this model clearly underestimates the average pressure both in the cluster core and in the outskirts. We have fitted the data with a model of the form given in Eqn \ref{eqn:pressure} allowing $\alpha$, $\beta$ and $\gamma$ to vary as free parameters, and we obtain $\alpha=0.73\pm0.07$, $\beta=2.85\pm0.19$, and $\gamma=0.68\pm0.04$. This best-fit model is shown as a solid black line in the pressure panel of Figure \ref{deproj}, and also seems to under predict several of the data points beyond a radius of 500 kpc. We also note that, while the pressure profiles along different arms show a remarkable agreement out to 500~kpc, their dispersion suddenly increases in the outskirts (marked by a grey background in the figure).

\begin{figure*}
\begin{center}
\includegraphics[width=0.45\textwidth]{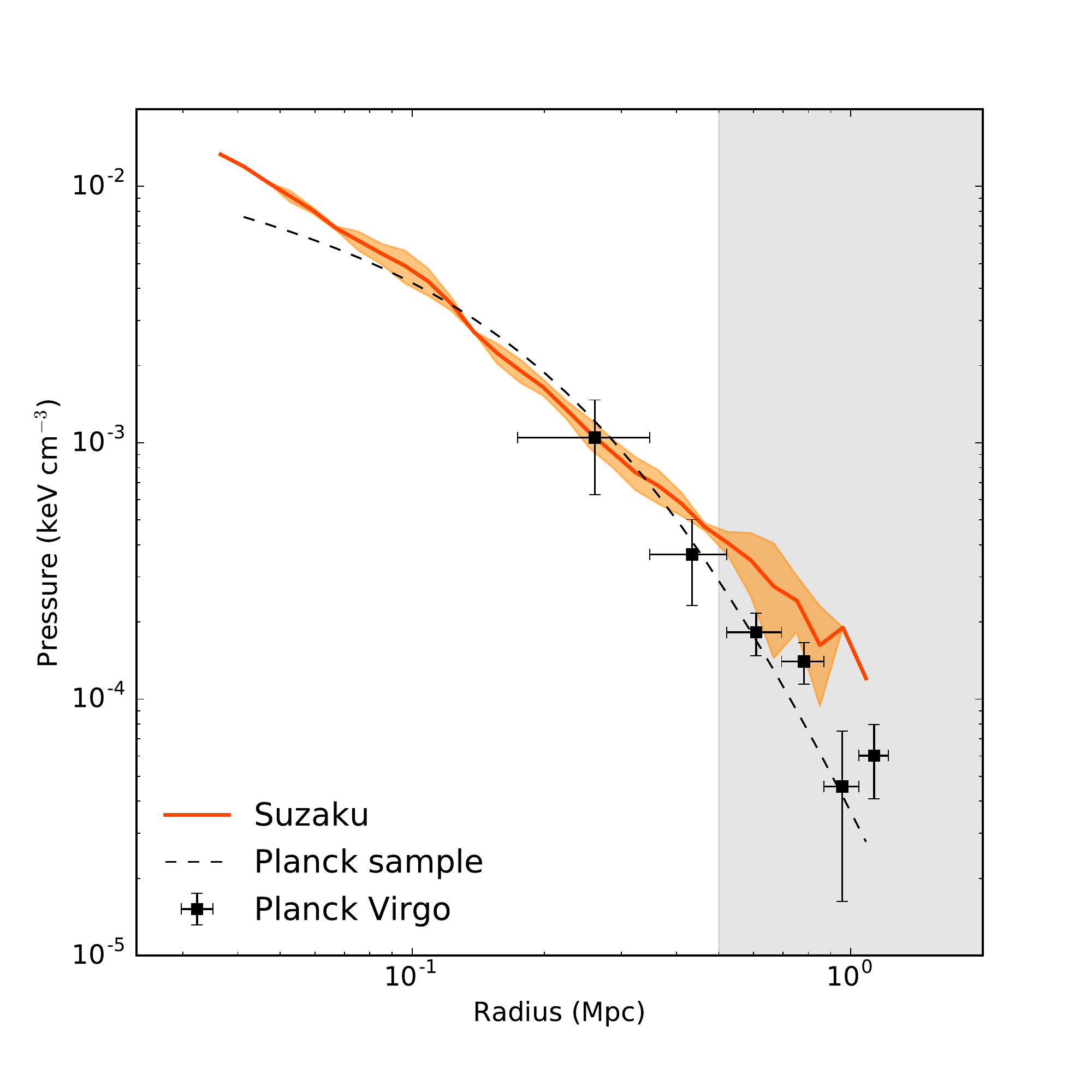}
\includegraphics[width=0.45\textwidth]{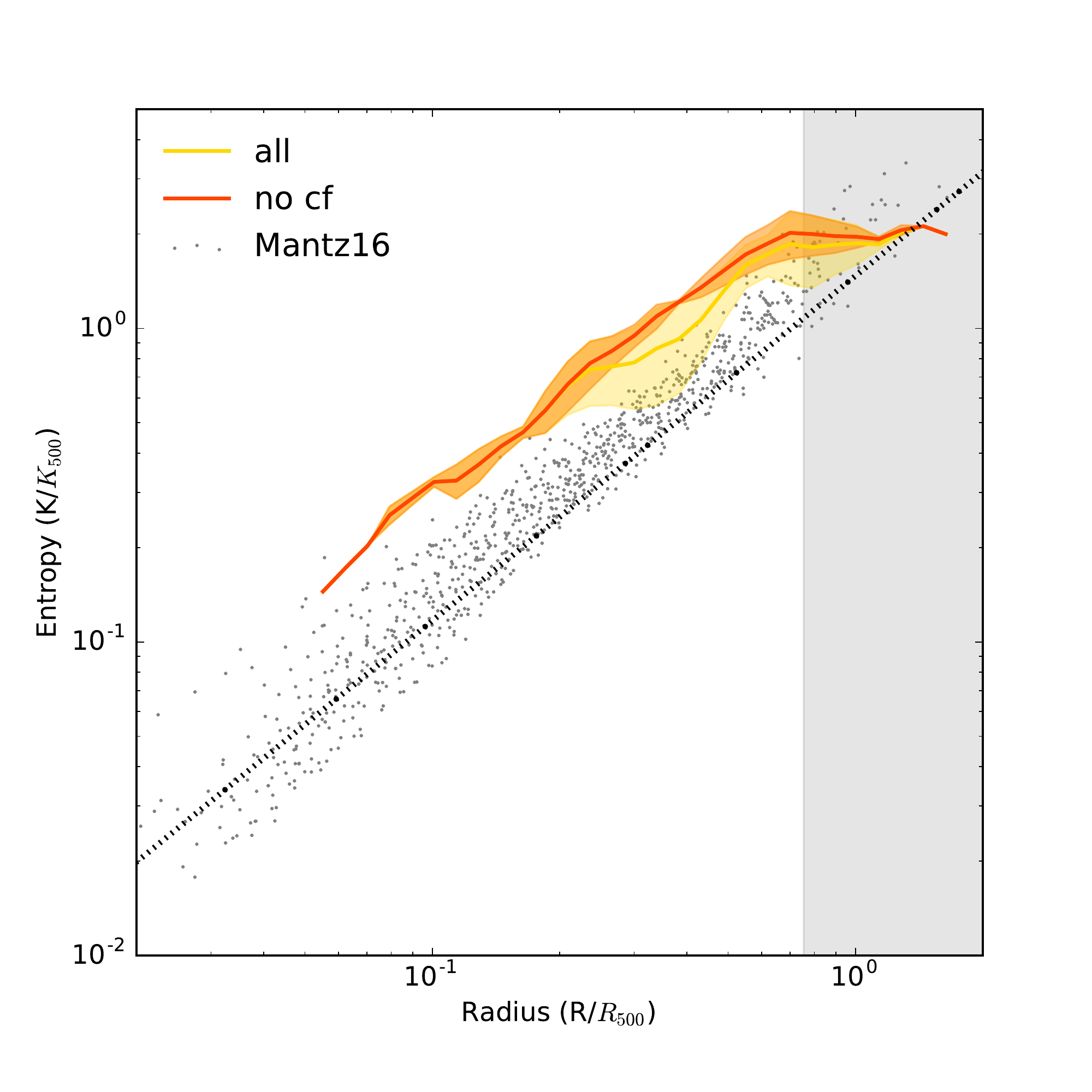}
\caption{{\it Left:} Comparison between the azimuthally averaged pressure profile measured in this work (shown in dark orange; all data points including the cold fronts are used) and the results reported by \citet{planck_virgo} based on Planck SZ data (shown as black data points). For reference, the average profile from a sample of 62 clusters observed by \citet{planck2013} is shown as a dashed line. 
{\it Right:} Azimuthally averaged entropy profiles obtained using all data points (yellow) and excluding the S, W, and large-scale N cold fronts (dark orange). The scaled entropy profiles for a sample of 40 massive, relaxed clusters observed with Chandra \citep{mantz2016thermo} are over plotted as grey dots. The expected entropy baseline profile given by $K/K_{500}=1.47(r/r_{500})^{1.1}$ is shown as a dotted line.
The area beyond 500~kpc, corresponding to the observed entropy flattening and increase in dispersion between the pressure profiles along different arms, is marked with a grey background in both panels.}\label{planckcomp}
\end{center}
\end{figure*}

In Fig. \ref{planckcomp}, we compare the pressure measurements obtained from Suzaku and from Planck SZ observations of the Virgo Cluster. The radial profiles along the four azimuths covered by Suzaku were interpolated onto a common set of radii in order to calculate the azimuthally averaged pressure profile, shown as a dark orange line. We then reevaluated this average for all of the 4-combinations based on the set of four arms (allowing for repetition) and calculated the standard deviation of the results, represented as the width of the band in the plot. This roughly reflects the uncertainty in the average profile that can be expected due to azimuthal asymmetry (note that statistical uncertainties are not included; the width of the band decreases at the outermost radii because only one or two arms can still be used to evaluate the variance). To obtain the pressure from the Planck SZ data, we performed a deprojection of the azimuthally averaged Compton parameter $y_C$  shown in Fig. 13 of \citet{planck_virgo}, under the assumption of spherical symmetry. The results are displayed as black points in Fig. \ref{planckcomp}, showing only the radial range relevant for comparison to the Suzaku measurements (although the SZ data extend further out). To estimate the error bars on the Planck data points, we randomly generated 1000 $y_C$ profiles assuming a normal distribution with centroids and widths at each radius given by the best-fit values and error bars presented in \citet{planck_virgo}. We repeated the deprojection for each realisation, and calculated the standard deviation of the resulting pressure profiles. We note that the region corresponding to M49 seems to have been included in the analysis published by \citet{planck_virgo}, and an excess in $y_C$ can be seen at the location of this sub-halo (outermost data point in Fig. \ref{planckcomp} and several additional measurements beyond this radius which are not included in our plot). We verified that interpolating over that excess does not significantly change the deprojected profile at smaller radii: the data point at a radius of $\sim$1~Mpc increases by about $1\sigma$, while the effect on all other measurements inside this radius is negligible.

On average, in the region beyond 500~kpc, the azimuthally averaged pressure determined from the SZ effect shows a systematic deficit compared to the azimuthally averaged Suzaku values, and we observe an increase in dispersion between the X-ray derived pressure profiles along different arms (illustrated by the width of the orange band in Figure \ref{planckcomp}).

\subsection{Entropy profiles}

In the central parts of the cluster, the entropy profiles increase monotonically with radius along all the four azimuths probed by Suzaku, the only exceptions being the southern and western cold fronts which appear as prominent dips. Around a radius of 450--550~kpc, the E and N arms show a prominent deficit compared to the S and W directions; the region of low entropy along the E arm is localised and most probably associated with the halo of M90, while the N arm shows a broader dip extending over several hundred kpc. Importantly, beyond 500~kpc, in the same region of the cluster where we observe an increase in dispersion between the pressure profiles along different arms, as well as an excess in the average X-ray pressure compared to Planck SZ data, the entropy profiles along all azimuths flatten out and the steep increase with radius seen in the central parts disappears. This region is marked by a grey background in Figure \ref{deproj}.

The flattening of the observed entropy profiles in cluster outskirts, spatially coincident with an excess in X-ray pressure compared to SZ measurements, has been reported previously for a number of systems \citep[e.g.][]{Walker13,urban2014}, and can be explained by the onset of gas clumping that results in an overestimation of the ICM density from X-ray measurements, affecting both the entropy and pressure profiles simultaneously. \citet{walker2012c} propose an analytical function that empirically describes the observed entropy profiles in the $r>0.2r_{200}$ range for a sample of systems observed with Suzaku. This function has the form:

\begin{equation}
K/K\left(0.3r_{200}\right)=A\left(r/r_{200}\right)^{1.1}\exp\left[-\left(r/Br_{200}\right)^2\right],
\label{eqn:entropy1}
\end{equation}
with $\left(A,B\right)=\left(4.4_{-0.1}^{+0.3},1.00_{-0.06}^{+0.03}\right)$. This model is shown with a solid line in the entropy panel of Figure \ref{deproj} (no fitting was performed other than scaling by $K\left(0.3r_{200}\right)$). It provides a relatively good description of the data, indicating that the shape of the entropy profile in Virgo is similar to previous observations of other clusters. 

To further illustrate this entropy flattening, we computed the azimuthally averaged profile and its variance in the same way as the mean pressure profile shown in Figure \ref{planckcomp}. The average for all data points is shown in yellow, while the average obtained excluding the S, W, and large-scale N cold fronts and M90 halo are shown in dark orange. The figure shows, even more strikingly, the change in slope of the entropy profile at a radius of about 500~kpc.

While the shape of the entropy profile in Virgo is consistent with the onset of gas clumping, its normalisation differs significantly from theoretical predictions.
In the absence of any additional heating or cooling, shock heating during the gravitational collapse of massive clusters of galaxies is expected to result in an entropy profile of the form
\begin{equation}
\frac{K}{K_{500}}=1.47\left(\frac{r}{r_{500}}\right)^{1.1},
\label{eqn:pratt}
\end{equation}
where $K_{500}=106\,\text{keV}\,\text{cm}^2\left(\frac{M_{500}}{10^{14}M_{\odot}}\right)^{2/3}\left(\frac1{f_{\rm gas}}\right)^{2/3}E(z)^{-2/3}$ \citep{voit2005,pratt2010}, with $E(z)=\sqrt{\Omega_{\rm m}(1+z)^3+\Omega_\Lambda}$ denoting the ratio of the Hubble constant at redshift $z$ with its present value and $M_{500}$ the total cluster mass enclosed within $r_{500}$. For the parameters obtained from our mass modelling analysis ($r_{500}=662.6$~kpc and an azimuthally averaged enclosed $f_{\rm gas}$ at this radius of $\sim0.10$, see Section \ref{sect:mass}), this translates to the dotted line shown in the entropy panel of Figures \ref{deproj} and \ref{planckcomp}, which clearly under predicts the data at almost all radii. 

A natural explanation for the entropy excess in the Virgo Cluster centre compared to the predictions from the \citep{voit2005} baseline model is the presence of non-gravitational heating and gas redistribution due to either pre-heating or AGN feedback. Because of the shallower gravitational potential, this entropy excess is expected to be stronger in Virgo relative to more massive clusters. In Figure \ref{planckcomp} we also over plot the scaled entropy profiles for a sample of 40 massive, relaxed clusters observed with Chandra \citep{mantz2016thermo}, showing that this is indeed the case.

However, given that the entropy flattening and pressure excess start already at a radius of 500~kpc, which is inside the cluster's $r_{500}$, the overall normalization of the entropy profile compared to the \citep{voit2005} prediction is uncertain. If the $f_{\rm gas}$ value that we measure at $r_{500}$ is biased high compared to the real value due to gas clumping, the normalisation by $K_{500}$ (which contains a factor of $f_{\rm gas,500}^{-2/3}$) could also be wrong. In order for the function described by Eqn \ref{eqn:pratt} to match the average entropy measured in the Virgo Cluster at $r_{500}$, the enclosed value of $f_{\rm gas}$ at this radius would have to be $0.074\pm0.010$, which is indeed smaller than the measured average of $\sim0.10$. 

The most likely conclusion is that the observed profiles are a combination of non-gravitational heating and gas clumping happening simultaneously, making it difficult to disentangle the two effects. Note however that non-gravitational heating alone would not explain both the entropy flattening and X-ray pressure excess compared to SZ measurements. 
 
\section{Mass modeling}\label{sect:mass}

We estimate the distribution of gravitating mass in the Virgo Cluster using an adaptation of the \texttt{clmass} model described by \citet{clmass}, with the additional assumption that the gravitational potential follows an NFW shape \citep{NFW}. In short, the \texttt{nfwmass} model employed here acts as a mixing model in XSPEC, combining concentric spherical shells of isothermal gas assumed to be in hydrostatic equilibrium within an NFW potential. The resulting projected spectra are compared to the data and the NFW parameters are adjusted until a best-fit is found\footnote{The \texttt{nfwmass} model variant is also written by P. Nulsen, although it is not described in \citet{clmass}. We have modified it to allow on-the-fly adjustment of the beta model used to account for projected emission, to enforce consistency with the measured density profile, as discussed in \citet{mantz2014}. That work also includes a general overview of the \texttt{nfwmass} implementation.}. 

We modelled the four different azimuths probed by the Suzaku mosaic in parallel within a common gravitational potential but with the gas temperature and density values allowed to vary independently between the arms. As in Section \ref{sect_deproj}, we account for emission from gas that lies beyond the outermost annulus, and which is projected onto this annulus, using a beta-model extrapolation whose parameters are allowed to vary separately for each arm. Uncertainties in the parameters are determined using Markov chain Monte Carlo sampling\footnote{https://github.com/abmantz/lmc}. We obtain a best-fit NFW scale radius $r_s=111\pm3$~kpc and a concentration of $c=8.8\pm0.2$, implying $M_{500}=0.83\pm0.01\times10^{14}$~M$_\odot$, $r_{500}=662.6\pm3.4$~kpc, and $M_{200}=1.05\pm0.02\times10^{14}$~M$_\odot$, $r_{200}=974.1\pm5.7$~kpc. 

For the N, W, and S arms, the scatter in the concentration parameter among these three directions amounts to 23\%, while the scatter in $M_{200}$ is only 8.3\%. Unfortunately, the region inside the NFW scale radius along the E arm is poorly sampled by the placement of the mosaic pointings; fixing the scale radius and fitting only the normalization of the NFW profile for this arm, we obtain a value for $M_{200}$ that is within the scatter of the other three directions. Given the underlying assumption that the profile follows an NFW shape, we further note that the resulting best-fit parameters are robust to deviations from hydrostatic equilibrium and asymmetry in the cluster outskirts (since the fit is determined mostly by the inner regions with better signal-to-noise). Leaving out all the data beyond a radius of 400~kpc, which roughly corresponds to the distance of M86, the nearest clear sub-halo to the Virgo cluster centre, we obtain $c=8.9\pm0.4$ and $M_{200}=1.01\pm0.06\times10^{14}$~M$_\odot$, in good agreement with the parameters estimated from the full Suzaku mosaic.

\begin{figure}
\begin{center}
\includegraphics[width=0.45\textwidth]{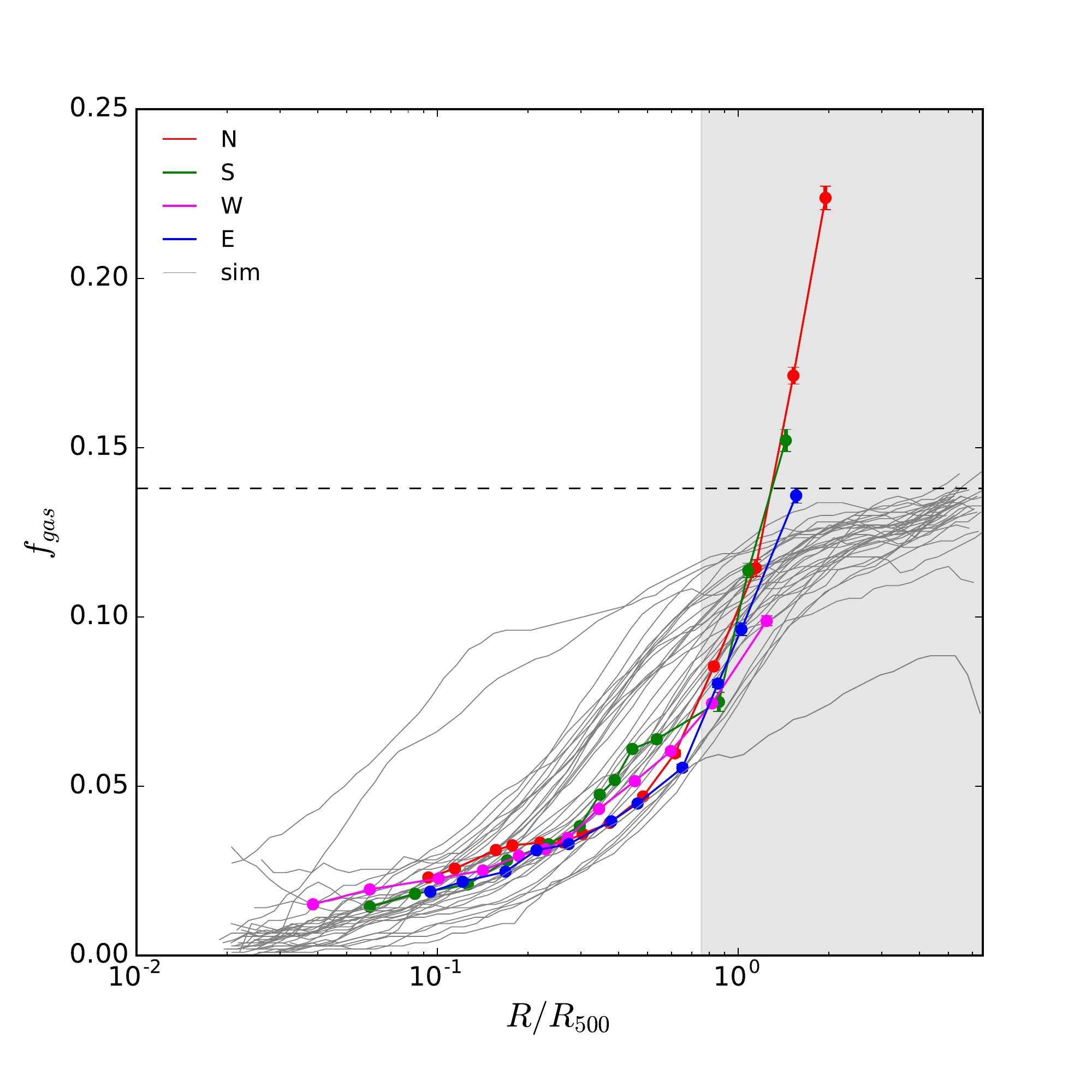}
\caption{Gas mass fraction profiles for the four different azimuths probed by Suzaku (in colour), compared to a sample of simulated clusters in the same mass range \citep{Rasia2015} shown as grey lines. The horizontal dashed line shows the expected gas mass fraction based on the universal baryon fraction reported by \citep{planck_cosmo2016}, and assuming that about 12\% of baryons are in stars. The region where we conclude that gas clumping begins to affect the entropy and pressure profiles is marked with a grey background.}\label{fig:fgas}
\end{center}
\end{figure}

This analysis also allows us to calculate the gas mass fraction along each of the four Suzaku arms, shown in Figure \ref{fig:fgas}. For a universal baryon fraction of 0.157 \citep{planck_cosmo2016}, and assuming that about 12\% of baryons are in stars \citep{lin2004,gonzalez2007}, the cumulative gas mass fraction is expected to reach a value of about 0.138, shown as a horizontal dashed black line. However, the $f_{\rm gas}$ profiles along the cluster's major axis clearly exceed this value. 

Recently, \citet{Rasia2015} presented a set of cosmological hydrodynamical simulations of galaxy clusters where the effects of stellar and AGN feedback, together with artificial thermal diffusion, were implemented such that a dichotomy between cool core and non cool core clusters is successfully reproduced. These simulations also provide a prediction for the expected redistribution of gas in cluster cores, including the approximate effects of AGN feedback. In Figure \ref{fig:fgas}, we over-plot the gas mass fraction profiles for 34 clusters in this simulation, with masses comparable to that obtained for the Virgo Cluster ($M_{200}$ between 0.5--1.4$\times10^{14}$~M$_\odot$). The profiles have been rescaled to match the universal baryon fraction of 0.157 cited above.

Up to $\sim r_{500}$, the measured gas mass fraction profiles for all of the four azimuths in the Virgo Cluster are broadly consistent with the expectations from these numerical simulations. For the western arm, the measurements do not reach significantly further than $\sim r_{500}$. However along the other three arms, and in particular along the north-south long axis of the cluster, the $f_{\rm gas}$ profiles show a very steep increase at large radii that is in tension with the expectations from numerical models. \citet{Vazza13} used numerical simulations to study the impact of a limited azimuthal coverage, concluding that the estimate of the enclosed baryon fraction can be biased from the true value by $\pm20$ per cent in systems with large-scale asymmetries (even in the absence of clumping). This can contribute to the $f_{\rm gas}$ excess observed, although along the northern arm the measured value exceeds the expectation by a much higher factor of 60\%. In combination with the pressure excess compared to the SZ measurements and the entropy flattening described in the previous sections, this is consistent with the onset of gas clumping in the cluster outskirts, with the clumping factor likely higher along the cluster's major axis. 


\section{Two-temperature spectral models}\label{sect_2T}

\begin{figure*}
\begin{center}
\includegraphics[width=0.45\textwidth]{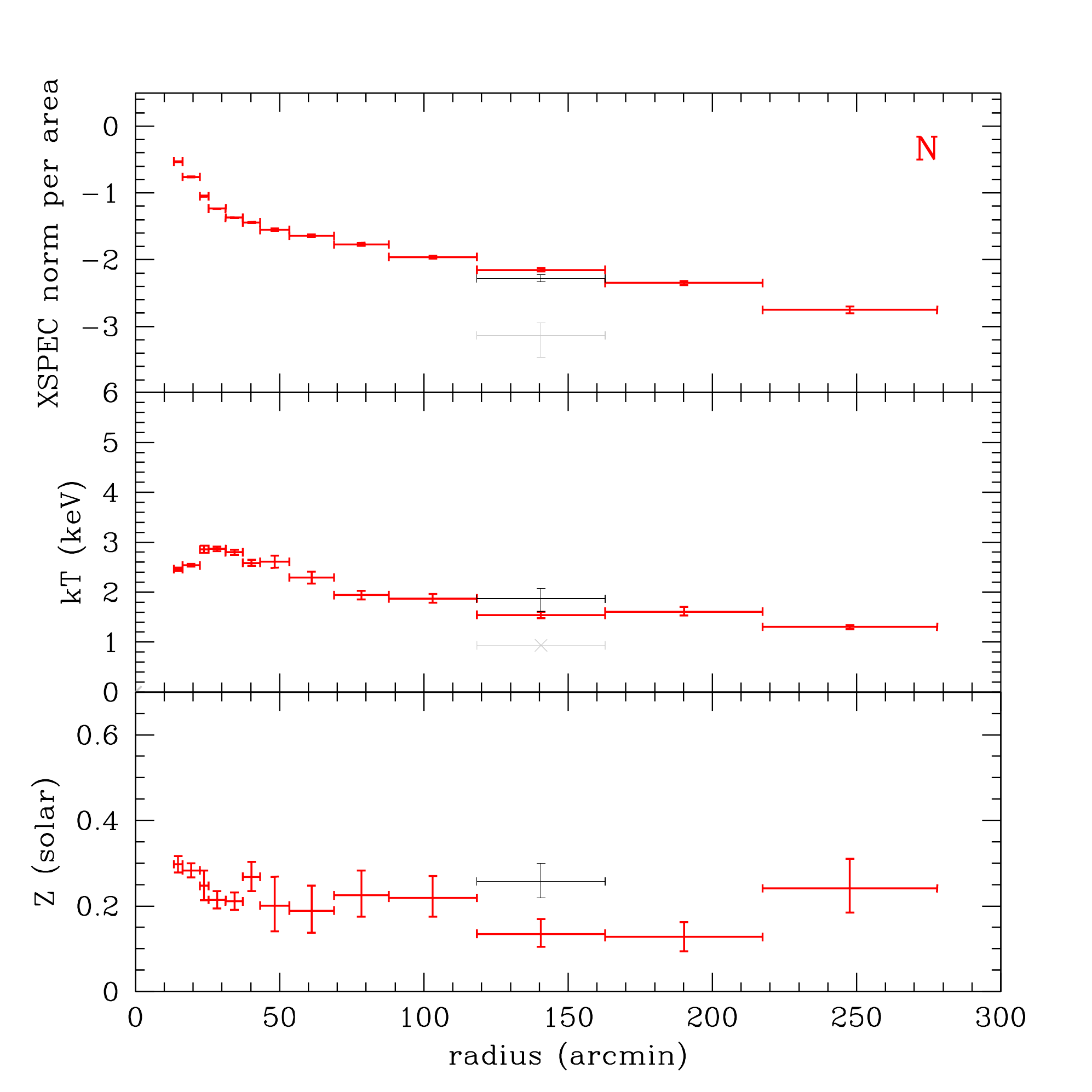}
\includegraphics[width=0.45\textwidth]{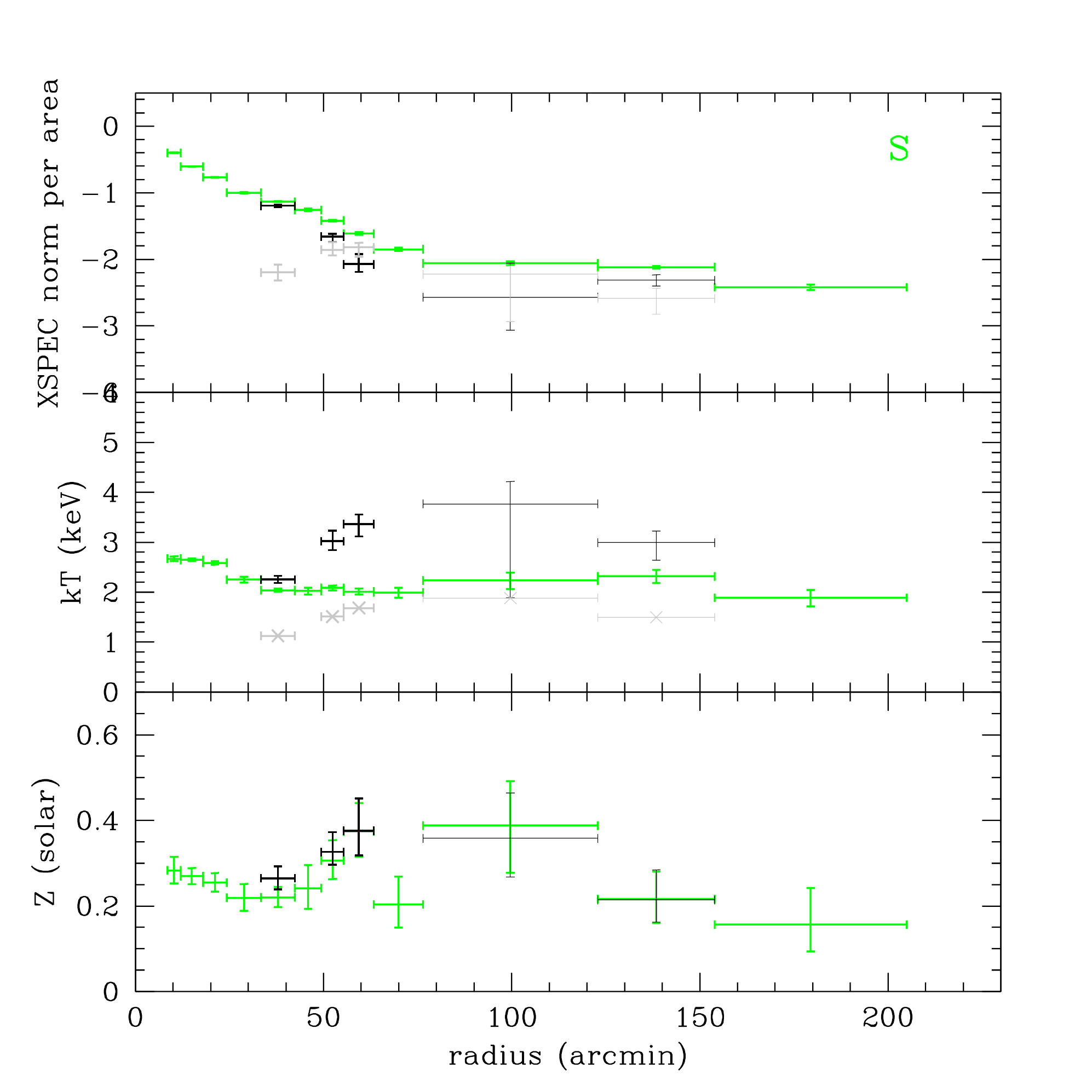}
\includegraphics[width=0.45\textwidth]{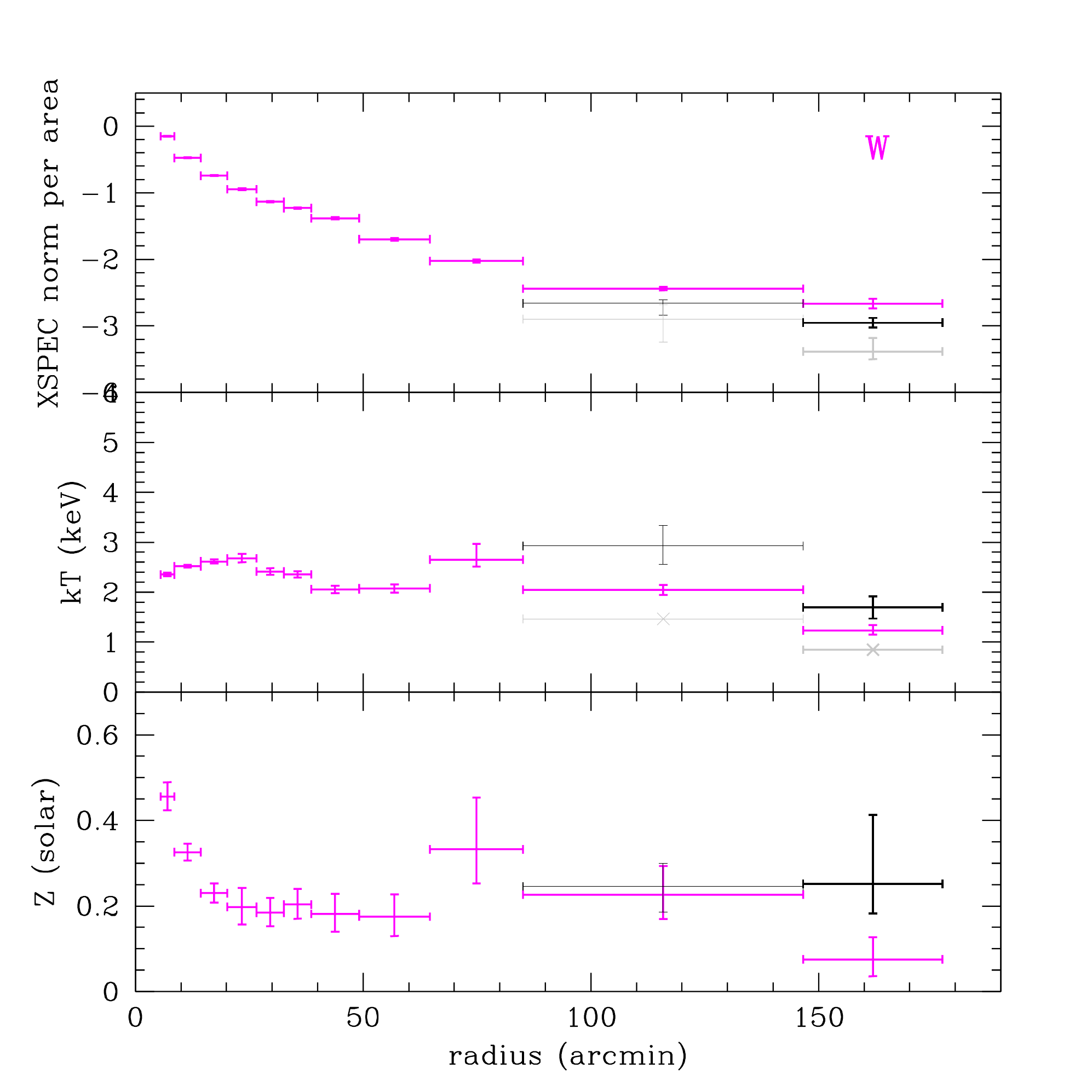}
\includegraphics[width=0.45\textwidth]{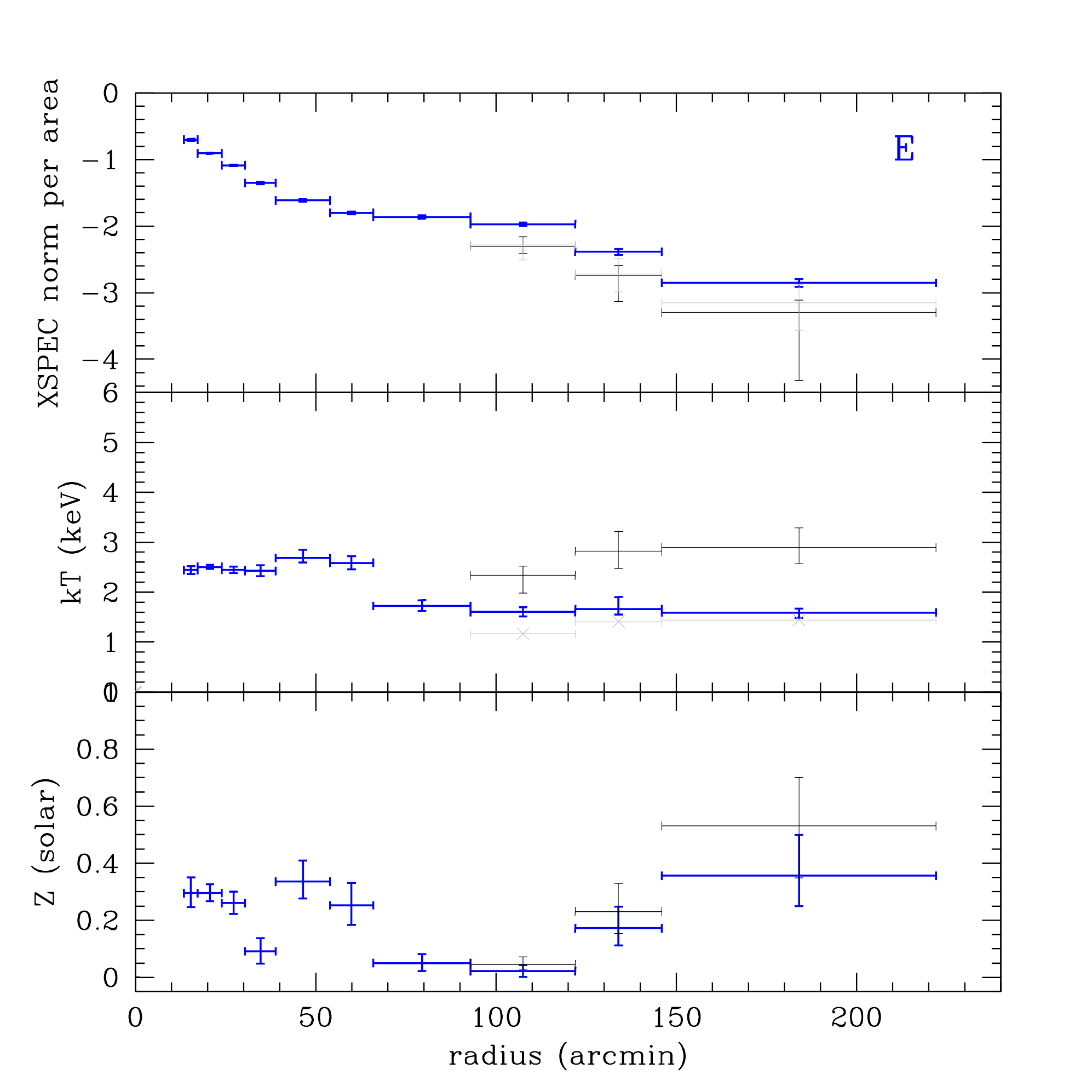}
\caption{The effect of two-temperature spectral models on the radial profiles of the projected spectrum normalisation, temperature, and metallicity along each of the four azimuths. In black, we show the temperature and spectrum normalisation of the hotter component obtained from the two-temperature fits, as well as the metallicity (assumed to be the same between the two gas phases). The temperature and spectrum normalisation of the cooler component are shown in grey. No error bars are plotted on the temperature of the cooler component, since this parameter was fixed to one half of the temperature of the hotter gas. Up to a radius of 100 arcmin, we only show the results for annuli where the second component was detected with a significance above $3\sigma$, while above 100 arcmin we include all detections above $1\sigma$ significance. Results for annuli where the significance was above $3\sigma$ are plotted with a thick line, while those with significance between 1--3$\sigma$ are shown with a thin line. For comparison, single temperature fits are shown in colour.
}\label{2Th}
\end{center}
\end{figure*}

In order to investigate thermodynamical substructure in the ICM beyond the single temperature approximation, we have also employed a spectral model consisting of two thermal components. Since constraining this model requires a larger number of counts, we performed these spectral fits using the wide radial bins used in the deprojection analysis (Section \ref{sect_deproj}). In addition, to further improve the stability of the fit, the metal abundances of both thermal components were assumed to be the same, and the temperature of the cooler component was assumed to be a fixed factor of two lower than the temperature of the hotter gas. The results from this analysis are shown in Figure \ref{2Th}.

Beyond a radius of 100 arcmin (465~kpc), a two-temperature model is required with a significance above $3\sigma$ in only one of the considered annuli, namely the outermost region along the western arm (680--824~kpc). Here, the best fit model contains a hotter component with a temperature of $1.7\pm0.2$ keV, contributing about 73\% of the emission, and a cooler 0.85~keV gas phase contributing the remaining 27\% of the emission. In seven other annuli located beyond 100 arcmin, a second temperature component is found at between 1--3$\sigma$ significance, while three regions (the outermost two northern and one southern annuli) do not require any additional thermal model beyond the single temperature approximation. 

In the higher signal to background regions between 20--100 arcmin, we only show the results where a two-temperature model was required with a significance above $3\sigma$. Only three such regions are found, and all appear to be associated with the southern cold front region. 

At radii less than 20 arcmin, the innermost annuli along the east and west, as well as the second and third annuli in the southern profile also show the presence of a two-temperature component. However, we suspect that this is most probably due to scattered light from the bright core of M87 and therefore we did not include these measurements in Figure \ref{2Th}.

Although the contribution of multi-phase gas appears to be small and the temperature and emissivity of most of the annuli considered here are not significantly affected, the overall best-fit metallicity does depend on the thermal model employed. Following the procedure described in Section \ref{sec_met}, upon excluding the annuli located near M87 and M49, as well as the southern cold front and the eastern arm that may be affected by instrumental systematics, we again fitted a constant model to the metal abundance measurements as a function of radius and obtain an average value of $0.22\pm0.01$ Solar, with an acceptable $\chi^2=22.4$ for 19 d.o.f. Importantly, a linear model of the form $Z(r)=Z_{0,2T}-r\times Z_{r,2T}$ now no longer provides a significant improvement, with $Z_{0,2T}=0.23\pm0.02$ Solar, $Z_{r,2T}=0.04\pm0.04$ Solar/Mpc, and $\chi^2=21.2$ for 18 d.o.f. 

Therefore, the radial gradient seen in Section \ref{sec_met} was artificially caused by multi-phase structure possibly related to the onset of gas clumping in the cluster outskirts; when this multi-temperature structure is accounted for, even in a naive way through a simple two-temperature fit, a uniform metal abundance distribution is recovered. However, at first glance, while the metallicity in Virgo is constant as a function of radius and azimuth, its mean, consistent with $\sim0.2$ Solar, is still lower than reported in the Perseus Cluster by \citet{werner2013nat}. 

We performed several additional tests to investigate the robustness of this conclusion. We focused on the annuli outside a radius of 300~kpc for all azimuths (which excludes the large-scale cold fronts), and omitted, as before, the M49 halo and the E arm. All remaining annuli were fit in parallel with a common metallicity, with the temperature and normalisation allowed to vary independently between the different regions. A single temperature fit yields an average metal abundance of $0.19\pm0.02$ Solar, fully consistent with the best-fit constant model reported in Section \ref{sec_met}. 

A two temperature model (with the temperature of the cooler component set to one half of that of the hotter component) yields an abundance of $0.26\pm0.02$ Solar, slightly higher than the average value of $0.22\pm0.01$ reported by fitting a constant to the radial profiles obtained from the two-temperature fits in Figure \ref{2Th}. In other words, when we assume the metallicity to be the same between all annuli, the 2T spectral fit prefers to adjust the normalisations of the cool gas to yield a slightly higher abundance than if the annuli are fitted independently from each other. Averaging the Fe abundances outside 300~kpc reported by \citet{Simionescu2015}, we obtain a similar value of $0.26\pm0.02$ Solar for a 2T fit with a different assumption (the temperature of the colder gas was fixed to 1~keV rather than tied to one half of the temperature of the hotter phase) and including the E arm; using the SPEX atomic line data base \citep{kaastra1996} as opposed to {\it apec} in this case yields an even higher Fe abundance of $0.32\pm0.03$ Solar, in agreement with the Perseus measurements. 

In turn, a single temperature {\it apec} fit excluding the strongest Fe-L lines (in the energy band spanning 0.9--1.2 keV) that are known to be the most prone to biases related to the presence of multi-temperature yields a best-fit abundance (assuming Solar ratios for Mg, Si, S to Fe) of $0.31\pm0.05$ Solar. Therefore, given uncertainties in the atomic line databases and the exact multi-phase structure in the outskirts, we cannot confidently conclude that the average metal abundance in the outskirts of the Virgo Cluster is indeed lower than that measured in the Perseus Cluster. 

\section{Discussion} 

\subsection{Large-scale sloshing}\label{sect_sloshing}

One of the first important discoveries in the study of galaxy clusters with Chandra was that, in addition to shock fronts, another type of substructure could also generate remarkably sharp edges in the X-ray surface brightness maps of the ICM \citep{markevitch2000}. These features, now referred to as ``cold fronts'', consist of relatively dense, cool gas found on the inner (bright) side and low-density hot gas on the outer (faint) side of the sharp density discontinuity; the density and temperature gradients are opposite and compensate each other such that, unlike the case of shocks, the pressure across cold fronts is continuous \citep[for a review, see][]{markevitch2007}. The combination of low temperature and high density also translates into a significantly lower entropy in the regions located on the inside of cold fronts. 

Cold fronts in cooling cores are believed to be due to the sloshing (or, more accurately, swirling) of the gas in the gravitational potentials of clusters \citep{markevitch2001}. Numerical simulations show that sloshing can be induced easily by a minor merger, where a subcluster falls in with a nonzero impact parameter \citep{tittley2005,ascasibar2006}. The cool core of the X-ray halo of M87 is known to host two cold fronts at radii of about 33 kpc towards the south-east and 90 kpc towards the northwest \citep{simionescu2007,simionescu2010,Werner2016}. \citet{roediger2011} performed numerical simulations tailored to reproduce these two observed cold fronts and thereby deduce the initial trajectory, mass of the sub cluster, and age of the minor merger responsible for generating these features. These hydrodynamical models {\sl predict} the presence of asymmetries in the ICM properties extending much further out than the 90 kpc cold front.

In this work, we are able to confirm this prediction for the first time. We detect a surface brightness excess extending from 30--50 arcmin (140--233 kpc) along the western arm, as well as a surface brightness excess between 20--60 arcmin (90 -- 280 kpc) towards the south. The temperature gradients across these features support the cold front interpretation, with the gas on the fainter, outer side being hotter than the gas on the inner, denser side. As expected, the pressure profiles at 50 arcmin towards the west and 60 arcmin towards the south are continuous, while the entropy profiles show evident deficits associated with the gas on the inner side of these large-scale cold fronts (Fig. \ref{deproj}). Moreover, the detection of multi-temperature gas between 30--60 arcmin towards the south (Fig. \ref{2Th}) is also consistent with a cold front interpretation. In this region, a low-temperature gas parcel originating closer to the cluster core has likely been transported outward via gas sloshing; hotter plasma typical of the current radius of the cold front is located in projection in front of and behind this cooler gas parcel, giving rise to the multi-temperature structure.

All of the thermodynamical properties of the western and southern surface brightness excesses therefore suggest that these features are part of a very large-scale spiral pattern, a typical signature of gas sloshing. 

In addition, at even larger radii of around 450--550~kpc, the entropy profiles towards the N and E show prominent deficits compared to the S and W; while the region of low entropy along the E arm is most probably associated with the halo of M90, the N arm shows a broader dip extending over several hundred kpc, which may be due to an extension of the large-scale sloshing out to these large cluster-centric distances. In the XMM-Newton image \citep[see Fig. 2 of ][]{simionescu2010}, the inner part of the sloshing spiral appears to run counter-clockwise, with the inner cold front at 33~kpc along the SE connecting to the 90~kpc cold front along the NW, which then links to a broad surface brightness excess seen along the southern part of the XMM-Newton mosaic and along the S arm with Suzaku. From the Suzaku observations, we now know that the outer edge of this over-dense, bright region corresponds to the southern cold front at 280~kpc. Beyond this, the spiral feature presumably further connects to the 233~kpc cold front towards the south-west (which lies closer to the centre due to the overall elongation of the spiral already seen in the XMM-Newton data). Continuing in a counterclockwise pattern, we expect the sloshing to then extend towards the W-NW and possibly reach the N arm creating the entropy dip around 450--550~kpc.

While observational and numerical efforts to understand the formation and physics of swirling cold fronts were initially directed at the bright inner cores of clusters, in recent years it has become evident that the signatures of gas sloshing extend out to much larger radii than previously thought. The present discovery of a large-scale spiral-like pattern in the Virgo Cluster joins a fast-growing list of similar features reported in the Perseus Cluster \citep{simionescu2012}, Abell 2142 \citep{rossetti2013}, Abell 2029 \citep{PaternoMahler2013}, and RXJ2014.8-2430 \citep{walker2014}. It is remarkable that these structures can remain coherent over such a broad range of spatial scales, and more detailed studies of the effects of hydrodynamical instabilities on these large-scale cold fronts may reveal important new details about ICM physics.

\subsection{Possible shocks near the virial radius}

Shock heating is primarily responsible for increasing the temperature of the diffuse ICM and allowing it to shine in X-ray light. This is one of the main aspects of baryonic physics that contributes to shaping large-scale structure in the present Universe. Generally speaking, large-scale shocks typically occur either at the very outer edges of otherwise relatively relaxed clusters of galaxies, at the boundary where lower temperature gas from the surrounding cosmic web filaments is entering their deeper gravitational potential wells (these are so-called ``accretion shocks''), or during mergers of two systems of comparable mass (``merger shocks''). Hydrodynamical simulations performed by \citet{molnar2009} suggest two distinct sets of accretion shocks: ``virial shocks'' with Mach numbers of 2.5-4, located at radii 0.9-1.3 times the virial radius, and stronger ``external'' shocks located farther out, at about three times the virial radius, with Mach numbers of order 100. Both types of shocks are predicted to cover about 50\% of the surface area at the given characteristic radii. 

While shocks associated with major cluster mergers have been found and studied in many systems \citep[for a far from exhaustive list of examples, see][]{markevitch2005,macario2011,russell2012,shimwell2014,uchida2016}, the X-ray emission at the typical location of accretion shocks is extremely faint, such that these features have yet to be detected unambiguously. Because shocks are thought to be responsible for accelerating particles up to the relativistic energies required to produce the synchrotron emission observed in the radio band \citep{ensslin1998}, a very efficient strategy to search for shock candidates in the X-ray faint outskirts of clusters (beyond about $r_{500}$) so far has been the use of follow-up observations of cluster radio relics \citep[e.g.][]{akamatsu2013,ogrean2013,ogrean2014b,vanWeeren2016}. Many of these studies reveal complications in the interpretation of the observational results, with the temperature and density jumps sometimes suggesting inconsistent shock Mach numbers \citep[e.g.][]{ogrean2014a}. In general, shocks in the cluster outskirts are seen more easily in the temperature than in the surface brightness or density profiles \citep[see radial profiles of the sample of radio relic clusters presented by][]{akamatsu2013}. This may arise from the fact that the density gradient is typically much steeper than the temperature gradient; in combination with the broad PSF of Suzaku, and non-spherical projection effects, this smears out density jumps seen against a steep underlying profile, while temperature jumps set against a flatter radial distribution are easier to identify.

Although located at the clusters' peripheries, shocks associated with radio relics still belong to the category of ``merger shocks''. To our knowledge, so far the only ``virial shock'' candidate published to date was found by \citet{urban2014}, located very close to $r_{200}$ along one of the eight azimuths in the Perseus Cluster probed by a previous Suzaku Key Project. \citet{urban2014} report shock Mach numbers of 1.43 from the density jump and 3.17 from the temperature jump at this candidate shock, once again showing that, at large radii, the contrast of such a feature is stronger in the temperature than the surface brightness profile. Additional possible variations of the degree of gas clumping further complicate the identification of the nature of this substructure. 

The thermodynamic profiles in the Virgo Cluster exhibit several additional intriguing features which could be indicative of the presence of shocks near the virial radius. Namely, we find two regions along the southern and western arms where the temperature is significantly enhanced. Along the southern arm, the temperature in the minimum surface brightness ``saddle'' point located 1~Mpc away from M87 and 233~kpc from M49 is enhanced by a factor of $1.70\pm0.34$ compared to neighbouring annuli. If this is a result of shock heating, then we are witnessing an early onset of a merger shock generated just as the M49 halo is crossing the virial radius of the Virgo Cluster. However, with the current data quality, we cannot rule out that the temperature increase is due simply to adiabatic compression. Along the western arm, we find an even larger temperature jump of a factor $3.14\pm0.71$ located 605~kpc from M87. Although this putative shock is located relatively close to the centre (about half of the Virgo Cluster's overall virial radius), its location is intriguing because the cluster is clearly squeezed from this direction. The surface brightness drops off much faster along the west compared to any of the other azimuths covered by Suzaku, such that the putative shock is located very close to the outer edge of the detection limit along this direction. The high temperature region we detect is thus likely related to the physical reason for the prominent western surface brightness asymmetry. If the enhanced temperature is the result of shock heating, it remains unclear whether this is a ``virial'' or ``merger'' shock, although no obvious perturber is present that would favour the ``merger'' shock scenario. Unfortunately, neither the southern nor the western temperature enhancements correspond to detected density jumps that would be required in order to identify these features as bona fide shocks. As mentioned above, this is not entirely unusual, as shocks in the outer parts of clusters appear to be easier identified in the temperature rather than surface brightness profiles. Nevertheless, a more in-depth study of these features would be needed in order to establish their nature unambiguously. 

\subsection{Behaviour of the entropy, pressure, and gas mass fraction profiles at large radii}

Arguably one of the most intriguing discoveries made by Suzaku and followed up with XMM-Newton, Chandra, and a combination of ROSAT archival data and Planck, has been an apparent flattening (or in some cases even turn-over) of the entropy profile at large radii, away from the predicted power-law shape expected in a cluster formed by gravitational collapse with no additional heating or cooling \citep[for early results, see e.g.][]{George09,Hoshino10,Simionescu11,walker2012c}.

The physical mechanism responsible for this entropy flattening is still a matter of debate; \citet{Simionescu11} presented evidence that a non-uniform density distribution in these regions (``gas clumping'') would bias high the gas density inferred from X-ray emissivity (which is proportional to $n_e^2$), resulting in an underestimate of the entropy and leading to the observed flattening. These findings were later corroborated by \citet{Walker13} and \citet{urban2014}, who combined the baseline entropy and pressure profiles given in Eqn. \ref{eqn:pressure} and \ref{eqn:pratt} to derive the expected baseline density and temperature profiles and compared these to the Suzaku measurements.  Assuming the true pressure follows that obtained for a sample of 62 clusters presented in \citet{planck2013} (the SZ signal depends linearly on $n_e$ and is thus less sensitive to clumping bias), \citet{Walker13} and \citet{urban2014} show that the measured density typically lies above the baseline model, while the temperature within $r_{200}$ is in good agreement with the expected profile. This would suggest that a density overestimate, rather than a temperature underestimate, is responsible for the observed entropy flattening. Evidence for the onset of gas clumping has also been obtained from Chandra and XMM-Newton, whose better spatial resolution in principle allows for a direct imaging of the denser gas clumps. Using a very deep, 2~Ms Chandra X-ray Visionary Program observation of Abell 133, \citet{Morandi14} report a gas clumping factor that increases with radius, reaching $\sim2-3$ at 0.9$r_{200}$ (of similar magnitude to the clumping factors inferred from Suzaku data). More recently, \citet{Tchernin16} show differences between the XMM-Newton X-ray surface brightness profiles in the outer parts of Abell 2142 obtained from the median versus the mean value of pixels within a given annulus; while the mean X-ray surface brightness leads to an overestimate of the density and flattening of the entropy profile, the median value which is much less sensitive to gas clumps recovers entropy values closer to the predicted power-law shape of \citet{voit2005}.

Meanwhile, \citet{Okabe14} argue that the universality of the scaled entropy profiles indicates that the thermalization mechanism is controlled by gravitation in a common way for all clusters, although the heating efficiency in the outskirts needs to be modified from the standard power-law, and that the flattening of the entropy profiles is caused by the steepening of the temperature, rather than the flattening of the gas density. They report a good agreement between the Suzaku X-ray pressure in the outskirts of Hydra A, Abell 478, Abell 1689, and the expected baseline pressure model from \citet{planck2013}, while the density overestimate due to gas clumping should have resulted in the X-ray pressure being higher than the SZ predictions. Several possible physical mechanisms that would reduce the thermalisation efficiency in the cluster outskirts have been proposed. \citet{Burns10} and \citet{Avestruz16} suggest that non-thermal gas motions driven by mergers and accretion play a significant role; \citet{Hoshino10} and \citet{Avestruz15} argue that the electron and ion temperatures in the outskirts may not have had time to equalise following the passage through the virial shock; \citet{Lapi10} propose that, as the clusters get older, the gas falls through a progressively smaller potential drop as the accretion shock moves out, reducing the entropy gain at the shock; \citet{Fujita13} suggest that the missing thermal energy in the outskirts is converted to cosmic ray acceleration. It is of course possible and perhaps very probable that all of the above mechanisms contribute to some extent to shaping the thermodynamical properties in the cluster outskirts. 

The results for the Virgo Cluster presented here show that the entropy profiles along all the probed azimuths level out beyond a radius of about 500~kpc, or 0.5$r_{200}$, in agreement with the initial conclusions reported by \citet{urban2011} using XMM-Newton observations of the N arm. At the same time, the pressure profiles at large radii show an excess, on average, compared to the \citet{planck2013} sample (Figure \ref{deproj}) and also compared to the Planck measurements performed specifically for this target \citep[][Figure \ref{planckcomp}]{planck_virgo}. Along the cluster's major axis running north-south, the measured enclosed baryon fraction clearly exceeds the cosmic mean at large radii. 

It is therefore very likely that the non-uniform distribution of the gas density in the Virgo Cluster plays a major role in the observed deviations from the expected self-similar entropy and pressure profiles. 
However, at the same time, the overall normalisation of the entropy profile is also affected by non-gravitational heating and gas redistribution due to AGN feedback and/or preheating, making the exact level of gas clumping difficult to evaluate quantitatively.



The four different azimuths in the Suzaku mosaic were chosen so as to cover both relaxed and disturbed directions and both the long and short axis of the cluster. If our present observations reflect a representative mean of the Virgo Cluster as a whole, then the excess in the azimuthally averaged X-ray pressure compared to the Planck SZ measurements for this system indicates that microscopic clumping at large radii is present and could affect all of the directions probed; this is indeed in line with the fact that the entropy is seen to flatten beyond a radius of 500~kpc along all arms. However, undoubtedly, ``macroscopic clumping'' in the form of a large-scale azimuthal asymmetry also influences the results. The pressure excess compared to the expected model based on Eqn \ref{eqn:pressure} is much more pronounced along the N-S axis, the same directions where we also measure a cumulative gas mass fraction clearly exceeding the cosmic mean value. This is most likely because, in addition to any small-scale clumps that may be present along all arms, in the N-S direction we are probing a large-scale structure filament connecting to the outskirts of the Virgo Cluster. 

We note that the entropy profiles along the E and W arms flatten at roughly the same level as the N and S directions, despite the fact that the latter two arms show a higher total (micro+macroscopic) clumping. This could suggest that gas clumping may not be the sole mechanism that contributes to the entropy deficit. Along the cluster's major axis, the flattening may be due primarily to the density excess associated with micro/macroscopic clumping and the contribution from large-scale structure filaments, while along the short axis we may additionally witness a lower heating or mixing efficiency and therefore a lower mean gas temperature that also affects the entropy deficit (note that the outermost E and W temperatures are systematically lower at the corresponding radii than along the N and S; upper left panel of Fig. \ref{deproj}). Finally we note that multi-temperature gas is detected with more than $3\sigma$ significance in the outermost annulus along the W arm, and at a lower significance level in seven other regions beyond 100~arcmin, including all three outer annuli along the E arm. This supports the idea that microscopic gas clumps are present not only along the cluster's major axis, and that perhaps along the short axis of the cluster these clumps are further from thermal equilibrium with the rest of the ICM allowing them to be identified more easily as a separate spectral component. 


\subsection{A low-redshift, low-mass anchor for the concentration--mass relation}

The now standard $\Lambda$CDM cosmological model under the cold dark matter (CDM) paradigm, within which the majority of matter in the Universe is weakly interacting, has enjoyed great success in explaining astrophysical and cosmological data.
This model makes clear predictions for, among other things, the distribution of mass within the structures that form hierarchically in the Universe \citep[e.g.][]{NFW,bullock2001,gao2008}. In particular, dark-matter-only simulations suggest that concentrations for individual clusters grow steadily as they accrete through minor mergers. While sufficiently disruptive mergers can `reset' the concentration parameter to lower values (c $\sim$ 3), the overall trends largely reflect differences in the typical accretion histories of haloes as a function of mass and redshift \citep{ludlow2012, klypin2016}. Given that very few precise measurements of the concentration parameter exist for clusters with comparably low mass and redshift to Virgo, here we investigate how the addition of the Virgo Cluster data can improve the observational constraints on the concentration--mass relation and its evolution.

\begin{figure}
\begin{center}
\includegraphics[width=\columnwidth]{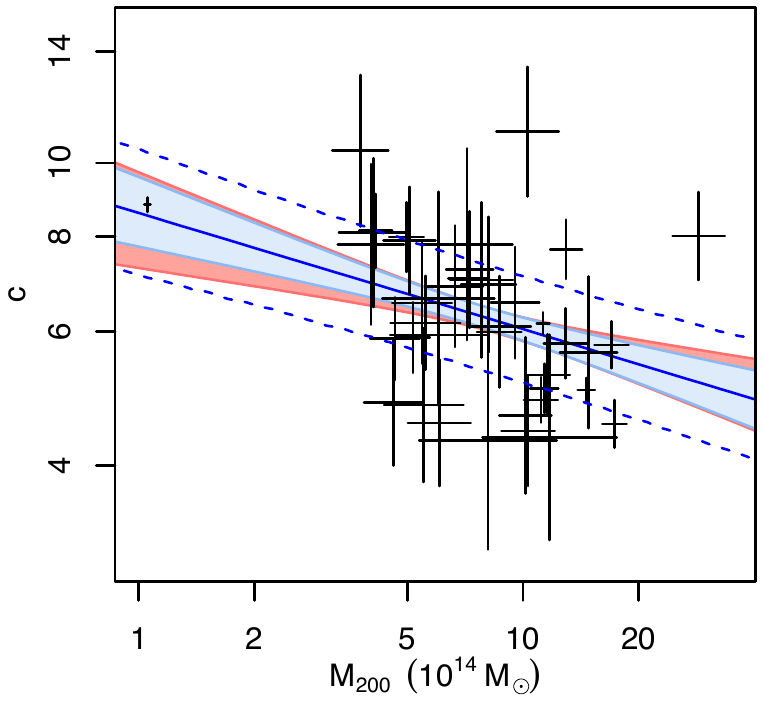}
\caption{NFW concentration parameter as a function of mass for the Virgo Cluster, compared to the sample of massive, dynamically relaxed clusters, measured from Chandra data \citep{mantz2016}. The $1\sigma$ confidence interval obtained from fitting the Chandra data alone is shown in red. The combined constraints from Chandra and the Virgo Suzaku data are shown with light blue shading. Including intrinsic scatter yields the 68.3 per cent predictive region delineated by dashed blue lines.}\label{cMfig}
\end{center}
\end{figure}

In Figure \ref{cMfig}, the measured parameters from the Virgo mass modelling analysis are compared with those of a sample of massive, dynamically relaxed
clusters, measured from Chandra data by \citet{mantz2016}. Following that work, we model concentration as a power law in mass and $1+z$, 
\begin{equation}
c=c_0 \left( \frac{1+z}{1.35} \right)^{\kappa_\zeta} \left( \frac{M_{200}}{10^{15}M_\odot}\right)^{\kappa_m},\label{cMeq}
\end{equation}
including a log-normal intrinsic scatter,  $\sigma_{\rm{ln} \:c}$, as a free parameter. We fit this model to the Chandra cluster data set and the additional Virgo data point using the method given by \citet{mantz2016}, which accounts for the strong anti-correlation between measurements of mass and concentration. The best-fitting parameters are $c_0= 6.2\pm0.5$, $\kappa_\zeta= -0.15\pm0.22$, $\kappa_m= -0.15\pm0.05$, and $\sigma_{\rm{ln} \:c}= 0.16\pm0.03$. The best fit is shown in the figure, along with its 68.3 per cent predictive intervals (including the intrinsic scatter). These results are in excellent agreement with those of \citet{mantz2016}, but with significantly reduced uncertainty in the power-law indices: respectively 29 per cent and 13 per cent in the exponents controlling the mass and redshift dependence of the concentration. As pointed out by \citet{mantz2016}, these values of $\kappa_\zeta$ and $\kappa_m$ are also in good agreement with measurements from gravitational lensing \citep[e.g.][]{merten2015}. The value of $\kappa_m$ moreover remains in good agreement with predictions from simulations \citep{bullock2001,gao2008} for low-redshift systems, even as the addition of a new measurement at $\sim5$ times lower mass than the Chandra sample has greatly reduced its uncertainty. The redshift dependence is less strong than predicted by numerical simulations ($\kappa_\zeta\sim-1$), most likely due to the fact that the Chandra cluster sample is composed of relaxed systems only, while an equivalent cut on dynamical state is not performed for the N-body simulation results.

We note that, although the Virgo Cluster is clearly unrelaxed, unlike the clusters included in the Chandra sample, the mass and concentration are determined robustly: as pointed out in Section \ref{sect:mass}, the values measured from the entire Suzaku mosaic are in very good agreement with those obtained leaving out all the data beyond the radius of M86, the nearest sub-halo from the centre of the cluster. The intrinsic scatter that we include when computing the best-fit concentration-mass relation in Figure \ref{cMfig} largely accommodates the scatter in concentration of $\sim23$\% between individual arms in Virgo. Moreover, although cross-calibration uncertainties between Suzaku and Chandra may affect the best-fit temperature (and therefore total mass), these are typically important for hot systems and are much less prominent for low-temperature clusters such as Virgo. Therefore, combining the Virgo Cluster data point with the measurements obtained for the Chandra sample should yield reliable results.

\subsection{Implications for the early chemical enrichment scenario}\label{sect_chemdisc}

Based on the uniform Fe abundance distribution on large scales in the Perseus Cluster, \citet{werner2013nat} conclude that most of the metal enrichment of the intergalactic medium occurred before the entropy in the ICM became stratified, probably during the period of maximal star formation and black hole activity at redshifts $z=2-3$. This early enrichment and mixing is expected to produce a constant abundance pattern on large scales in all massive clusters; the Virgo Cluster data set offers us the opportunity to test this prediction on smaller physical spatial scales (due to the proximity of this target), and for a less massive system. 

Using the present data set, \citet{Simionescu2015} already reported a uniform distribution of the Si/Fe, S/Fe, and Mg/Fe metal abundance ratios in the ICM using relatively coarse extraction regions. The constant chemical composition throughout the ICM of the Virgo Cluster further supports the early enrichment model. Here, assuming a constant alpha-element to Fe ratio, we present metallicity profiles with a finer radial binning. The extraction regions from which the single-temperature fit results are reported have typical sizes of 20$\times$(10--30) arcmin (between 90$\times$45 and 90$\times$140~kpc), while two-temperature models are obtained from regions that are roughly twice larger ($90\times180\pm75$~kpc). 

Along the N, W, and S arms, apart from known substructures which can be expected to have a different metallicity (the haloes of M87 and M49 and the southern large-scale sloshing front), we generally recover a remarkably uniform metal abundance distribution.
While the metallicity profiles obtained from a single-temperature modelling of the ICM suggest a decrease with radius, this trend can be (at least in part) caused by multi-phase structure related to the onset of gas clumping in the cluster outskirts. When this multi-temperature structure is accounted for, even in a naive way through a simple two-temperature {\it apec} fit, we find that a constant model provides a good fit to the distribution of measured metal abundance values versus radius, with an average of $0.22\pm0.01$ Solar and an acceptable $\chi^2=22.4$ for 19 d.o.f., and no significant radial gradient is detected.
A uniform metal abundance distribution is thus recovered on physical spatial scales roughly five times smaller than in the Perseus Cluster \citep[about $20\times10$ arcmin, corresponding to $400\times200$~kpc reported by][compared to $90\times180$~kpc in this work]{werner2013nat}.

An interesting exception from a uniform metal abundance profile is the eastern arm where, although on average the metallicity is consistent with the other azimuths, we observe strong fluctuations around this mean value. It is unfortunately unclear whether these fluctuations are physical or the result of a deteriorating gain stability or increased noise of the XIS detectors between AO-7 (when all other arms were observed) and AO-8.

Lastly, we note \citep[as already pointed out by][]{Simionescu2015} that the average abundance in the outer parts of the Virgo Cluster seems at first glance to be lower compared to the value of $0.29\pm0.01$ determined for the Perseus Cluster. This could in principle provide an interesting angle on the early enrichment scenario, which suggests that all massive clusters should have a common metal abundance at large radii. If the metallicity in the outskirts of the Virgo Cluster were indeed significantly lower than the Perseus Cluster, this could indicate that, during the early phase of cluster formation, the gravitational potential of the Virgo Cluster was too shallow to retain all metals produced at redshifts of 2--3, and conversely that the AGN and stellar feedback during that time were vigorous enough to expel metals from a protocluster that would evolve into a $10^{14}M_\odot$ system in the present Universe. 

However, as we point out in Section \ref{sect_2T}, even a weak multi-temperature structure of the ICM can influence the metallicity fits. The average metal abundance beyond a radius of 300~kpc becomes $0.26\pm0.02$ of the Solar value for a 2T {\it apec} model if all annuli are fit in parallel with a single metallicity. This in better agreement, although still smaller than the value measured in the Perseus Cluster. On the other hand, \citet{Simionescu2015} reported an even higher Fe abundance of $0.32\pm0.03$ Solar for the same region using a 2T SPEX fit, as opposed to {\it apec}.
For a single temperature fit excluding the strongest lines in the Fe-L complex that are known to be the most prone to biases related to the presence of multi-temperature, an {\it apec} model gives a best-fit average abundance in the outskirts of the Virgo Cluster of $0.31\pm0.05$, in good agreement with the Perseus Cluster measurements, where the metallicity measurements are based mainly on the strength of the Fe-K lines. 

\section{Conclusions}

We have presented the results from a Suzaku Key Project dedicated to mapping the outskirts of the Virgo Cluster in X-rays along several azimuths using more than 60 different pointings and over a megasecond of total net exposure time. This data set allowed us to perform detailed spectroscopic measurements of the thermodynamical properties in the intracluster medium of the nearest galaxy cluster out to its virial radius. Our conclusions can be summarised as follows:
\begin{itemize}
\item{The entropy profiles along all the probed azimuths increase with radius in the centre and flatten out beyond about 500~kpc, or 0.5$r_{200}$. The average Suzaku pressure at these large radii shows an excess compared to the Planck SZ measurements performed both for a sample of clusters and specifically for this target. In the same radial range, the dispersion between the pressure profiles for different arms also increases significantly, while the measured enclosed baryon fraction along the cluster's major axis exceeds the cosmic mean in the outermost annuli. We conclude that both ``microscopic clumping" in the form of small fluctuations which could be present along all probed azimuths and ``macroscopic clumping'' in the form of a candidate large-scale structure filament running along the north-south direction likely cause the observed deviations from the expected self-similar entropy and pressure profiles. However, the observed profiles are affected by a combination of non-gravitational heating and gas clumping happening simultaneously, making it difficult to disentangle the two effects and evaluate the exact level of gas clumping quantitatively.}
\item{The Virgo Cluster hosts a large scale pattern typical of gas sloshing, with the previously known cold fronts at 33~kpc along the SE and 90~kpc along the NW complemented by similar features at 280~kpc along the southern direction and 233~kpc along the western arm, the latter two which are identified for the first time in this work. The sloshing may further extend along a counterclockwise, highly elliptical spiral pattern and cause the broad entropy dip seen around 450--550~kpc along the northern arm.}
\item{We find two high-temperature regions located at radii of 1~Mpc towards the south and 605~kpc towards the west. We suggest that these are shocks associated with the ongoing formation of the cluster, although we could not identify the expected density jumps corresponding to the Mach numbers indicated by the measured temperature discontinuities. The southern high temperature region corresponds to the low surface brightness saddle point connecting the M87 and M49 haloes and is therefore most probably related to the ongoing interaction between these two subclusters.}
\item{We estimated the mass of the Virgo Cluster, assuming the ICM to be in hydrostatic equilibrium within a gravitational potential required to follow an NFW shape. We obtain a virial radius $r_{200}=974.1\pm5.7$~kpc (corresponding to $M_{200}=1.05\pm0.02\times10^{14}$~M$_\odot$) 
and a concentration $c=8.8\pm0.2$, consistent with previous expectations taking into account the dependence of the halo concentration on mass and redshift.
Combining the Virgo Cluster data point with existing measurements for a sample of massive, relaxed clusters observed with Chandra allows us to significantly reduce the uncertainties in the power-law exponents controlling the mass and redshift dependence of the concentration.}
\item{The metallicity profiles along the northern, western, and southern directions are generally consistent with a uniform metal distribution on scales as small as $\sim 90\times180$~kpc, supporting the early enrichment scenario in which most chemical elements were dispersed into the ICM by galactic winds at redshifts of $z=2-3$. Although multiphase structure and differences in the plasma modeling of the Fe-L complex make it difficult to determine the absolute metal abundance precisely, within systematic uncertainties, the average metallicity of the gas in the outskirts of the Virgo Cluster is consistent with the Z$\sim$0.3 solar measured in the Suzaku Key Project study of the Perseus Cluster. 
}
\end{itemize}

\section*{Acknowledgments}

We are grateful to J.M. Diego for providing the Compton Y-parameter profile for the Virgo Cluster from Planck data, to E. Rasia for providing the gas mass fraction profiles from a sample of simulated clusters, and to M. Takada for fruitful discussions. This work was supported in part by NASA
grant NNX13AI49G and by the US Department of Energy under contract number DE-AC02-76SF00515, as well as by the Lend\"ulet LP2016-11 grant awarded by the Hungarian Academy of Sciences.
The authors thank the Suzaku operation team and Guest Observer Facility, supported by JAXA and NASA. 


\bibliography{clusters}

\bibliographystyle{mnras}

\appendix

\section{Comparison with measurements from other satellites}

\begin{figure}
\begin{center}
\includegraphics[width=\columnwidth]{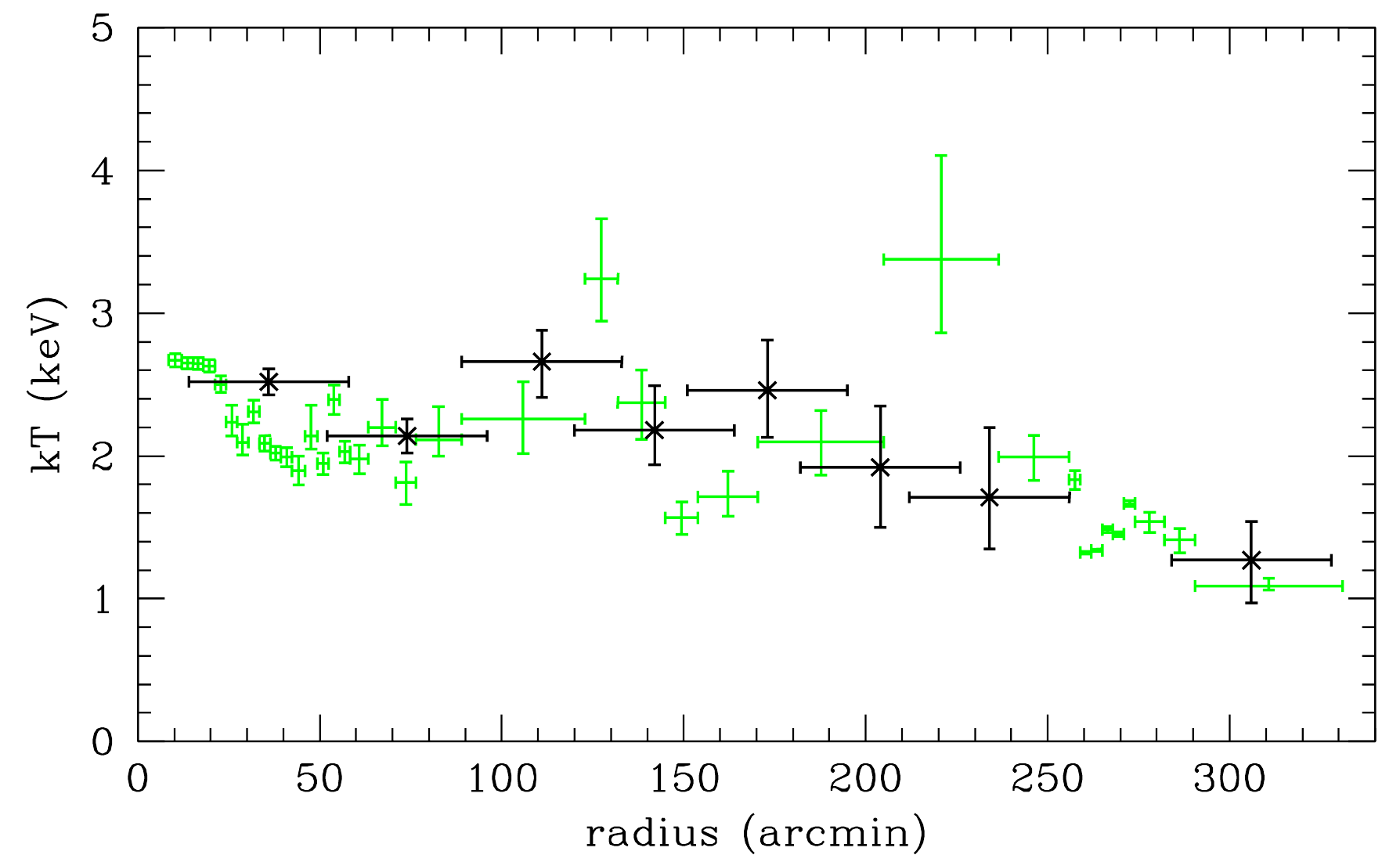}
\caption{Projected temperature profile along the southern arm from the Suzaku mosaic (shown in green) compared to previous results obtained with ASCA shown in black \citep{shibata2001}.}\label{ascacomp}
\end{center}
\end{figure}

\begin{figure}
\begin{center}
\includegraphics[width=\columnwidth]{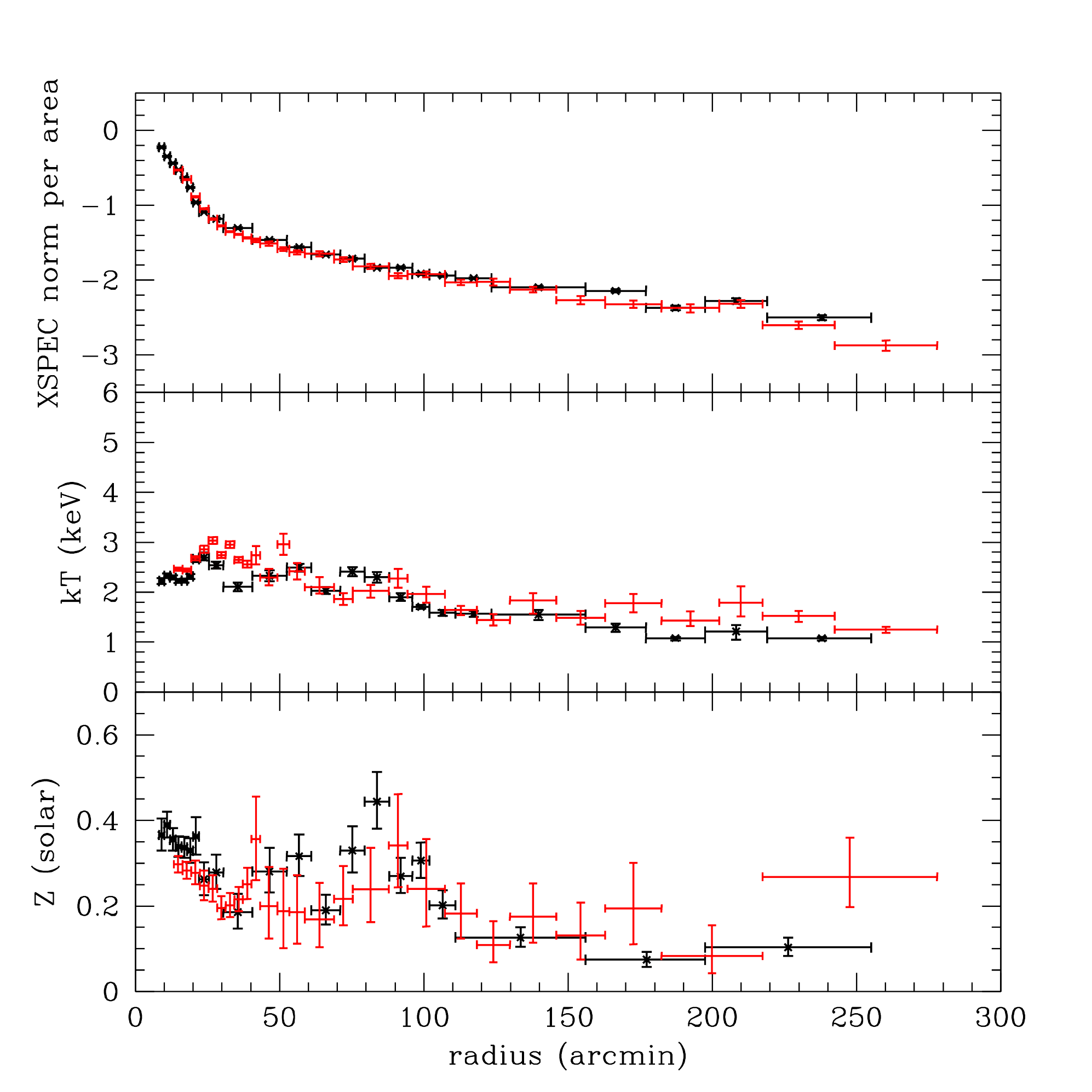}
\caption{Projected temperature, metallicity, and emission measure for the northern arm from XMM-Newton \citep{urban2011}
shown in black, compared to the Suzaku measurements for the same azimuth obtained from this work, shown in red.}\label{xmmcomp}
\end{center}
\end{figure}

\begin{figure*}
\begin{center}
\includegraphics[width=0.45\textwidth]{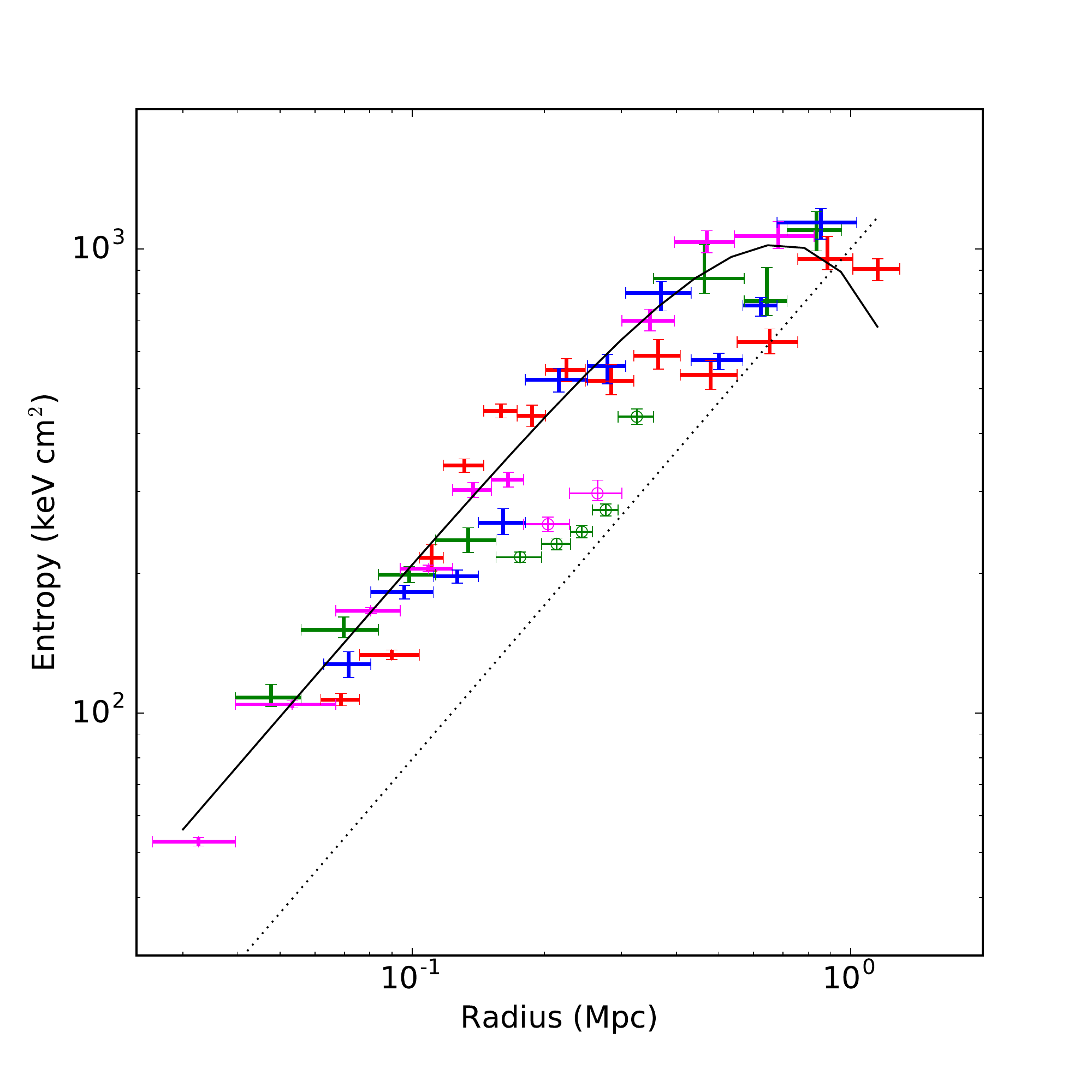}
\includegraphics[width=0.45\textwidth]{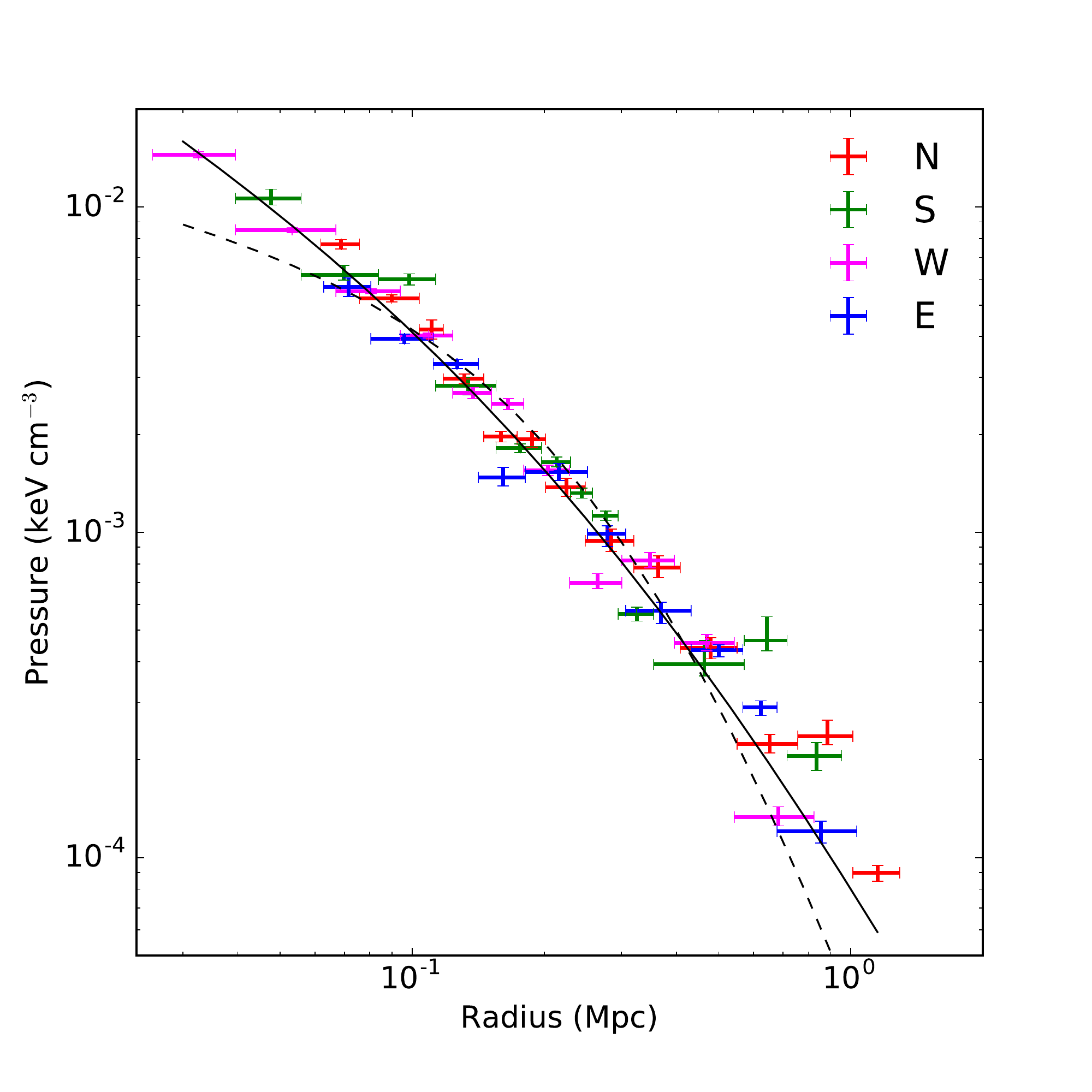}
\caption{The entropy (left) and pressure (right) profiles obtained when including a hot foreground model component. The black curves are identical to those in Figure \ref{deproj}.}\label{hf4}
\end{center}
\end{figure*}

The first studies of the temperature structure of the Virgo Cluster out to large radii were published by \citet{shibata2001} based on hardness ratio values from ASCA observations. The primary region of overlap between the ASCA pointings and the Suzaku mosaic used in this work is along the southern part of the Virgo Cluster, in the direction connecting the M87 and M49 haloes. In Figure \ref{ascacomp}, we have chosen the ASCA pointings that best overlap spatially with the Suzaku southern arm, and we compare the temperature profiles measured with these two different data sets. In general, we find an excellent agreement between the two temperature profiles, while the spatial resolution of the Suzaku measurements is clearly superior. This allows us to discover the region of high-temperature gas situated around 220 arcmin south of the cluster core, in the minimum surface brightness ``saddle'' point between M87 and the X-ray peak associated with the M49 subgroup; this feature was probably not visible with ASCA due to the much larger spatial binning. 

In addition, \citet{urban2011} studied the spectroscopically derived properties in the outskirts of the Virgo Cluster using a mosaic of XMM-Newton pointings extending towards the north of M87 out to beyond the cluster's virial radius. In Figure \ref{xmmcomp}, we compare the best-fit spectral normalisations, temperatures, and metal abundances measured with XMM-Newton to the results obtained from the northern arm with Suzaku. In general, we find a good agreement between the two data sets, although beyond $r_{500}$, the temperatures and metallicities derived with XMM-Newton seem to be systematically lower than measured by Suzaku, which may indicate some small residual systematic uncertainties likely associated with subtracting the instrumental background of XMM-Newton.

\section{Systematic uncertainties due to the potential presence of a hot foreground component}\label{sect_hf}

\citet{Yoshino09} report the detection of a thermal plasma with typical temperatures between 0.5--1.0 keV in some blank-sky observations. This component, which we will refer to as "hot foreground" (HF), is thought to be associated with emission from unresolved faint stars \citep{Masui09} and appears predominantly at low Galactic latitudes; it should therefore not contaminate the Virgo Cluster observations presented here. However, \citet{sasaki2016} report a significant HF component in their background modelling of the Virgo Cluster. We suspect that this is much more likely to represent low-level emission coming from beyond the Virgo Cluster's virial radius, rather than a hot Galactic foreground emission. Nonetheless, since the spectra do not allow us to distinguish between these two possibilities, here we discuss the potential systematic effects associated with including a HF component in the background model.

\begin{table}
\caption{Background model parameters for a fit including an HF component. The errors are statistical at 68\% confidence, except for the CXB normalisation for which we quote the standard deviation among the 11 Suzaku fields of view considered here.}
\begin{center}
\begin{tabular}{ccccc}
\hline
\hline
 & kT (keV) & arm & Suzaku norm \\
 & or $\Gamma$ &  & $\times 10^{-7}$/arcmin$^2$ \\
\hline 
CXB & 1.50 & all & $8.6\pm0.9$ (stddev) \\ 
\hline
 & & N & $7.4\pm$1.6 \\
GH & $ 0.20$ & W & $10.1\pm1.9$ \\
 & & S  & $11.8\pm3.3$ \\
 & & E &  $10.0\pm3.1$\\
\hline
 & & N & $0.79\pm0.20$ \\
HF & $0.82^{+0.09}_{-0.05}$ & W & $0.39\pm0.25 $  \\
 &  & S & $1.76\pm0.37 $  \\
 & & E & $1.58\pm0.39$ & \\
\hline 
\end{tabular}
\end{center}
\label{tab_hf}
\end{table}

We have fitted the Suzaku background fields as described in Section \ref{sect_bkgmo} with a model that now includes an additional HF component whose temperature is assumed to be the same between all the background spectra, while its flux is allowed to vary among the different azimuths. The results are shown in Table \ref{tab_hf}. The deprojected pressure and entropy profiles obtained with this new background model (including the beta correction described in Section \ref{sect_deproj}) are presented in Figure \ref{hf4}, together with the same model curves shown in Figure \ref{deproj} in the main body of the paper (these reference curves have not been re-fitted to match the new data points). All of our scientific conclusions are robust to this new background treatment.

\section{The list of Suzaku observations}
\label{obslist}

Table \ref{tab:observations} provides a list of all the Suzaku observations used in this work, including the coordinates of the aim-point, exposure time, and average Galactic column density.

\begin{table*}
  \caption{Suzaku observations used in this work.}
  \label{tab:observations}
  \scriptsize
  \centering
  \begin{tabular}{lccccccc}
  \hline
    Target & ObsID & Obs. date & RA & Dec & Exposure  & $n_H$ \\
     & &  &  &  &(ks) &(10$^{20}$ cm$^{-2}$) \\  
    \hline
C0       & ae801038010 & 2006-11-29 &  12 30 56.642 & +12 26 47.46  & 90 & 1.94   \\               
M49   & ae801064010 & 2006-12-03 &  12 29 46.053 & +8 00 14.94  & 90 & 1.53   \\   
\hline
N0 & ae803070010 & 2008-06-08 & 12 30 47.814 & +12 42 02.12 & 22 & 1.86 \\
N05 & ae803069010 & 2008-06-08 & 12 30 47.526 & +12 56 22.88 & 60 & 1.96 \\
\hline
N1  & ae807106010 & 2012-06-19 & 12 30 46.875 & +13 15 40.28 & 9 & 2.15 \\
N2  & ae807107010 & 2012-06-20 & 12 30 46.255 & +13 33 56.62 & 8 & 2.36 \\
N3  & ae807108010 & 2012-06-20 & 12 30 46.530 & +13 53 20.50 & 12 & 2.48 \\
N4  & ae807109010 & 2012-06-26 & 12 30 49.498 & +14 13 00.45 & 13 & 2.58 \\
N5  & ae807110010 & 2012-06-20 & 12 30 46.396 & +14 31 25.76 & 16 & 2.68 \\
N6  & ae807111010 & 2012-06-26 & 12 30 47.737 & +14 51 30.32 & 19 & 2.57 \\
N7  & ae807112010 & 2012-06-26 & 12 30 48.593 & +15 11 16.67 & 16 & 2.42 \\
N8  & ae807113010 & 2012-06-27 & 12 30 48.759 & +15 29 42.95 & 19 & 2.27 \\
N9  & ae807114010 & 2012-06-28 & 12 30 47.632 & +15 49 11.80 & 19 & 2.23 \\
\hline
N10 & ae806060010 & 2011-12-10 & 12 30 51.056 & +16 05 39.76 & 21 & 2.20 \\
N11 & ae806061010 & 2011-12-10 & 12 30 51.200 & +16 21 48.16 & 17 & 2.15 \\
N12  & ae806062010 & 2011-12-11 & 12 30 51.104 & +16 37 40.00 & 21 & 2.14 \\
N13 & ae806063010 & 2011-12-20 & 12 30 51.656 & +16 53 58.84 & 24 & 2.08 \\
N14 & ae806064010 & 2011-12-23 & 12 30 51.560 & +17 09 49.60 & 34 & 2.04 \\
N15 & ae806065010 & 2011-12-24 & 12 30 51.608 & +17 25 49.00 & 33 & 1.99 \\
N16 & ae806066010 & 2011-12-25 & 12 30 51.752 & +17 41 56.32 & 46 & 2.04 \\
\hline
W1 & ae807094010 & 2012-07-02 & 12 30 01.592 & +12 14 29.33 & 10 & 1.94 \\
W2 & ae807095010 & 2012-07-02 & 12 28 55.665 & +12 05 14.93 & 12 & 1.96 \\
W3 & ae807096010 & 2012-12-12 & 12 27 50.171 & +11 56 38.22 & 8 & 2.01 \\
W4 & ae807097010 & 2012-12-12 & 12 26 44.051 & +11 47 07.98 & 14 & 2.09 \\
W5 & ae807098010 & 2012-12-12 & 12 25 33.683 & +11 36 49.50 & 17 & 2.25 \\
W6 & ae807099010 & 2012-12-13 & 12 24 23.982 & +11 26 51.72 & 19 & 2.40 \\
W7 & ae807100010 & 2012-12-13 & 12 23 17.710 & +11 15 31.53 & 23 & 2.47 \\
W8 & ae807101010 & 2012-12-14 & 12 22 12.070 & +11 05 03.69 & 18 & 2.41 \\
W9 & ae807102010 & 2012-12-14 & 12 21 06.622 & +10 55 43.89 & 26 & 2.28 \\
W10 & ae807103010 & 2012-12-15 & 12 20 01.246 & +10 46 28.83 & 25 & 2.08 \\
W11 & ae807104010 & 2012-12-15 & 12 18 55.990 & +10 37 10.11 & 21 & 1.97 \\
W12 & ae807105010 & 2012-12-16 & 12 17 57.934 & +10 27 19.35 & 26 & 1.90 \\
\hline
S1 & ae807115010 & 2012-07-03 & 12 31 02.788 & +12 06 39.92 & 13 & 2.07 \\
S2 & ae807116010 & 2012-07-03 & 12 30 53.093 & +11 46 59.48 & 12 & 2.20 \\
S3 & ae807117010 & 2012-07-03 & 12 30 48.031 & +11 28 25.64 & 13 & 2.27 \\
S4 & ae807118010 & 2012-07-04 & 12 30 39.508 & +11 10 07.82 & 13 & 2.30 \\
S5 & ae807119010 & 2012-07-04 & 12 30 34.377 & +10 51 29.48 & 11 & 2.31 \\
S6 & ae807120010 & 2012-07-04 & 12 30 34.306 & +10 33 06.44 & 10 & 2.08 \\
S7 & ae807121010 & 2012-07-05 & 12 30 25.116 & +10 13 29.60 & 15 & 1.86 \\
S8 & ae807122010 & 2012-07-05 & 12 30 20.077 & +9 54 59.36 & 14 & 1.64 \\
S9 & ae807123010 & 2012-07-07 & 12 30 15.542 & +9 35 26.27 & 17 & 1.55 \\
S10 & ae807124010 & 2012-12-08 & 12 30 12.753 & +9 16 26.37 & 12 & 1.44 \\
S11 & ae807125010 & 2012-12-08 & 12 30 07.593 & +8 56 37.29 & 17 & 1.41 \\
S12 & ae807126010 & 2012-12-09 & 12 29 58.891 & +8 37 20.22 & 15 & 1.41 \\
S13 & ae807127010 & 2012-12-09 & 12 29 53.985 & +8 18 44.97 & 11 & 1.43 \\
S14 & ae807128010 & 2012-12-09 & 12 29 44.818 & +7 40 58.05 & 15 & 1.61 \\
S15 & ae807129010 & 2012-12-10 & 12 29 35.133 & +7 21 14.22 & 16 & 1.68 \\
S16 & ae807130010 & 2012-12-10 & 12 29 31.306 & +7 03 12.93 & 20 & 1.65 \\
S17 & ae807131010 & 2012-12-11 & 12 29 26.530 & +6 44 44.49 & 17 & 1.64 \\
S18 & ae807132010 & 2012-12-11 & 12 29 21.706 & +6 25 15.93 & 18 & 1.62 \\
S19 & ae807133010 & 2012-12-11 & 12 29 17.554 & +6 07 07.29 & 19 & 1.65 \\
\hline    
E1   & ae808116010  & 2013-07-08  & 12 32 14.048  & +12 32 00.27  & 9  & 2.01  \\             
E2   &  ae808117010 & 2013-07-08 & 12 33 20.323 & +12 41 34.55 & 11 & 2.09 \\            
E3  & ae808118010   & 2013-07-08 & 12 34 26.815 & +12 50 23.67 & 10 & 2.36  \\     
E4   & ae808119010   &  2013-07-09 & 12 35 33.390 & +12 59 40.23 & 9 & 2.55  \\               
E5   & ae808120010   &  2013-07-09 & 12 36 39.965 & +13 10 59.19 & 12 & 2.79  \\                             
E6   & ae808121010 &   2013-07-09  & 12 37 46.565 & +13 21 09.75 & 11 & 2.84 \\                
E7   & ae808122010   &  2013-07-10 & 12 38 58.631 & +13 31 28.73 & 16  & 2.65 \\      
E8 & ae808123010   &  2013-07-10 & 12 40 09.795 & +13 41 41.67 & 16  & 2.34 \\  
E9 & ae808124010   &  2013-07-11 & 12 41 16.490 & +13 50 44.91 & 15  & 2.11 \\  
E10 & ae808125010   &  2013-07-11 & 12 42 23.257 & +13 59 47.07 & 19  & 1.93 \\  
E11 & ae808126010   &  2013-07-12 & 12 43 24.601 & +14 09 04.35 & 20  & 1.87 \\  
E12 & ae808127010   &  2013-07-12 & 12 44 31.608 & +14 18 07.59 & 17 & 1.91 \\  
E13 & ae808128010   &  2013-07-13 & 12 45 38.591 & +14 26 57.15 & 19  & 1.95 \\  
\hline

\end{tabular} 
\end{table*}

\end{document}